\documentclass[12pt]{JHEP3} 
\usepackage{epsfig,multicol}
\usepackage{citesort}
 \large\normalsize

\newcommand{\be}{\begin{equation}}
\newcommand{\ee}{\end{equation}}
\newcommand{\bea}{\begin{eqnarray}}
\newcommand{\eea}{\end{eqnarray}}
\def\lsim{\raise0.3ex\hbox{$\;<$\kern-0.75em\raise-1.1ex\hbox{$\sim\;$}}}
\def\gsim{\raise0.3ex\hbox{$\;>$\kern-0.75em\raise-1.1ex\hbox{$\sim\;$}}}
\def\Frac#1#2{\frac{\displaystyle{#1}}{\displaystyle{#2}}}
\def\no{\nonumber\\}

\def\susy{\mbox{\tiny SUSY}}
\def\sm{\mbox{\tiny SM}}
\def\susy{\mbox{\tiny SUSY}}
\def\sm{\mbox{\tiny SM}}
\def\pmkz{\pi^-\overline{K}^0}
\def\pzkm{\pi^0{K}^-}
\def\ppkm{\pi^+{K}^-}
\def\pzkz{\pi^0\overline{K}^0}
\def\Kbar{\overline{K}}

\def\ddLR{(\delta^d_{LR})_{23}}

\def\duLLtwo{(\delta^u_{LL})_{32}}

\def\duLRtwo{(\delta^u_{LR})_{32}}
\def\du#1#2{{\left(\delta^u_{#1}\right)_{#2}}}
\def\dd#1#2{{\left(\delta^d_{#1}\right)_{#2}}}
\newcommand{\W}{{\scriptscriptstyle W}}
\def\dofourfigs#1#2#3#4#5{\centerline{
\epsfxsize=#1\epsfig{file=#2, width=7.5cm,height=6.2cm
} \hspace{0.1cm} \hfil \epsfxsize=#1\epsfig{file=#3,  width=7.5cm,
height=6.2cm
}} \vspace{1.25cm} \centerline{ \epsfxsize=#1\epsfig{file=#4,
width=7.5cm,height=6.2cm
} \hspace{0.1cm} \hfil \epsfxsize=#1\epsfig{file=#5,  width=7.5cm,
height=6.2cm
}} }

\title{CP asymmetries and branching ratios of $B\to K \pi$ in
supersymmetric models}

\author{Shaaban Khalil\\
$(1)$ Department of Mathematics, German
University in Cairo-GUC, New Cairo city, EL Tagamoa EL Khames Egypt.\\

$(2)$ Department of Mathematics, Ain Shams University, Faculty of
Science,
Cairo, 11566, Egypt.\\

E-mail: \email{shaaban.khalil@guc.edu.eg}}

\abstract{We analyze the supersymmetric contributions to the
direct and mixing CP asymmetries and also to the branching ratios
of the $B\to K \pi$ decays in a model independent way. We consider
both gluino and chargino exchanges and emphasize that a large
gluino contribution is essential for saturating the direct and
mixing CP asymmetries. We also find that combined contributions
from the penguin diagrams with chargino and gluino in the loop
could lead to a possible solution for the branching ratios puzzle
and account for the results of $R_c$ and $R_n$ within $b \to s
\gamma$ constraints. When all relevant constraints are satisfied,
our result indicates that supersymmetry favors lower values of
$R_c$. Finally we study the correlations between the mixing CP
asymmetry $S_{K^0 \pi^0}$ and mixing CP asymmetries of the
processes $B\to \phi K$ and $B\to \eta' K$. We show that it is
quite possible for gluino exchanges to accommodate the results of
that observables }
\begin{document}
\section{Introduction}
Recently the BaBar and Belle collaborations have measured the CP
averaged branching ratios and the CP violating asymmetries of
$B\to K \pi$ decays \cite{Aubert:2004qm,giorgi,sakai}. These
results, in addition to those from the $B\to \phi K$ and $B\to
\eta' K$, offer an interesting avenue to understand the CP
violation and flavor mixing of the quark sector in the Standard
Model (SM).

In the SM, all CP violating observables should be explained by one
complex phase $\delta_{CKM}$ in the quark mixing matrix. The
effect of this phase has been observed in kaon system. In order to
account for the observed CP violation in this sector,
$\delta_{CKM}$ has to be of order one. With such a large value of
$\delta_{CKM}$, the experimental results of the CP asymmetry of
$B\to J/\psi K_S$ are consistent with the SM. However, the
experimental measurements of the CP asymmetries of $B\to \phi K$,
$B\to \eta' K$ and $B\to K \pi$ decays exhibit a possible
discrepancy from the SM predictions. Furthermore, it is well known
that the strength of the SM CP violation can not generate the
observed size of the baryon asymmetry of the universe, and new
source of CP violation beyond the $\delta_{CKM}$ is needed.

In supersymmetric extensions of the SM, there are additional sources
of CP violating phases and flavor mixings. It is also established
that the SUSY flavor dependent (off-diagonal) phases could be free
from the stringent electric dipole moment (EDM) constraints
\cite{Abel:2000hn}. These phases can easily provide an explanation
for the above mentioned anomalies in the CP asymmetries of $B\to
\phi K$ and $B\to \eta' K$
\cite{Gabrielli:2004yi,PhiKs_gluino,Khalil:2003bi}. We aim in this
article to prove that in this class of SUSY models, it is also
possible to accommodate the recent experimental results of $B\to K
\pi$ CP asymmetries and branching ratios.

\begin{table}[t]
\centering
\begin{tabular}{|c|c|c|c|}
\hline
Decay channel  & BR $\times 10^6$ & $A_{CP}$  & $S_f$ \\
 \hline
$\bar{K}^0\pi^- $  & $24.1\pm1.3$ & $-0.02\pm 0.034$ & $-$ \\
\hline
$ K^- \pi^0$   & $12.1 \pm 0.8$ & $0.04 \pm 0.04$ & $-$ \\
\hline
$ K^- \pi^+$  & $18.2 \pm 0.8$ & $-0.113 \pm 0.019$ & $-$\\
\hline
$\bar{K}^0 \pi^0 $ & $11.5\pm 1.0$ & $- 0.09\pm 0.14$ & $0.34\pm0.28$ \\
\hline
\end{tabular}
\caption{The current experimental results for the CP averaged
branching ratios and CP asymmetries of $B\to K \pi$ decays.}
\label{table2}
\end{table}

The latest experimental measurements for the four branching ratios
and the four CP asymmetries of $B\to K \pi$ \cite{Aubert:2004qm}
are given in Table 1. As can be seen from this table, the measured
value of the direct CP violation in $\bar{B}^0 \to K^- \pi^+$ is
$A^{CP}_{K^- \pi^+}= -0.113 \pm 0.019$ which corresponds to a $4.2
\sigma$ deviation from zero. While the measured value of
$A^{CP}_{K^+ \pi^0}$, which may also exhibit a large asymmetry, is
quite small. As we will see in the next section, it is very
difficult in the SM to get such different values for the CP
asymmetries.

Also from these results, one finds that the ratios $R_c$, $R_n$
and $R$ of $ B\to K \pi$ decays are given by
\begin{eqnarray} R_c
&=&2\left[\frac{BR(B^+\to K^+\pi^0)+BR(B^-\to K^- \pi^0)}
{BR(B^+\to K^0 \pi^+)+ BR(B^- \to \bar{K}^0 \pi^-)} \right]= 1.00 \pm 0.08,\label{Rcresult}\\
R_n & = & \frac{1}{2} \left[\frac{BR(B^0 \to K^+ \pi^-) +
BR(\bar{B}^0 \to K^- \pi^+)}{BR(B^0 \to K^0 \pi^0) + BR(\bar{B}^0
\to \bar{K}^0 \pi^0)} \right]= 0.79 \pm 0.08, \label{Rnresult}\\
R&=&\left[\frac{BR(B^0\to K^+\pi^-) + BR(\bar{B}^0 \to K^- \pi^+)}
{BR(B^+\to K^0 \pi^+)+ BR(B^- \to \bar{K}^0 \pi^-)}
\right]\frac{\tau_B^+} {\tau_{B^0}}= 0.82 \pm 0.06.
\label{Rresult}
\end{eqnarray}

In the SM the $R_c$ and $R_n$ ratios are approximately equal,
however, the experimental results in
Eqs.(\ref{Rcresult},\ref{Rnresult}) indicate to $2.4\sigma$
deviation from the SM prediction. On the other hand the quantity
$R$ is consistent with the SM value. Here
$\tau_B^+/\tau_{B^0}=1.089\pm 0.017$. These inconsistencies
between the $A^{CP}_{K\pi}$ and the $R_c-R_n$ measurements and the
SM results are known as $K \pi$ puzzles.

These puzzles have created a lot of interest and several research
work have been done to explain the experimental
data~\cite{Khalil:2004yb,kpi-puzzle}. It is tempting to conclude
that any new physics contributes to $B\to K\pi$ should include a
large electroweak penguin in order to explain these discrepancies.
In SUSY models, the $Z$ penguin diagrams with chargino exchange in
the loop contribute to the electroweak penguin significantly for a
light right handed stop mass. Also the subdominant color
suppressed electroweak penguin can be enhanced by the
electromagnetic penguin with chargino in the loop. Therefore, the
supersymmetric extension of the SM is an interesting candidate for
explaining the $K \pi$ puzzles.

It is worth mentioning that also new precision determinations of the
branching ratios and CP asymmetries of $B\to \pi \pi$ have been
recently reported \cite{giorgi,sakai}. However, the SUSY
contributions to $B\to \pi \pi$, at the quark level, is due to the
loop correction for the process $b\to d q \bar{q}$, while the SUSY
contribution to $B\to K \pi$ is due to the process $b\to s q
\bar{q}$. Therefore, these two contributions are in general
independent and SUSY could have significant effect to $B\to K \pi$
and accommodates the new result, while its contribution to $B\to \pi
\pi$ remains small. Thus we will focus here only on SUSY
contributions to $B\to K \pi$.

In this paper, we perform a detailed analysis of SUSY contributions
to the CP asymmetries and the branching ratios of $B\to K \pi$
processes. We emphasize that chargino contribution has the potential
to enhance the electroweak penguins and provides a natural solution
to the above discrepancies. However, this contribution alone is not
large enough to accommodate the experimental results and to solve
the $K\pi$ puzzles. We argue that the gluino contribution plays an
essential rule in explaining the recent measurements, specially the
results of the CP asymmetries. Recall that other supersymmetric
contributions like the neutralino and charged Higgs are generally
small and can be neglected. The charged Higgs contributions are only
relevant at a very large $\tan \beta$ and small charged Higgs mass.
Therefore, we are going to concentrate on the chargino and gluino
contributions only.

The paper is organized as follows. In section 2 we study the CP
asymmetries and the branching ratios of $B\to K \pi$ in the SM. We
show that within the SM the $K\pi$ puzzles can not be resolved. In
section 3 we analyze the supersymmetric contributions, namely the
gluino and chargino contributions, to $B\to K \pi$. We show that a
small value of the right-handed stop mass and a large mixing
between the second and the third generation in the up-squark mass
matrix are required to enhance the chargino $Z$-penguin. Also a
large value of $\tan \beta$ is necessary to increase the effect of
the chargino electromagnetic penguin.

Section 4 is devoted to the constraints on SUSY flavor structure
from the branching ratio of $b\to s \gamma$. New upper bounds on the
relevant mass insertions are derived in case of dominant gluino or
chargino contribution. A correlation between the mass insertions
$\dd{LR}{23}$ and $\du{LL}{32}$ is obtained when both gluino and
chargino exchanges are assumed to contribute significantly. In
section 5 the SUSY resolution for the $R_c-R_n$ puzzle is
considered. We show that it is very difficult to explain this puzzle
with a single mass insertion contribution. We emphasize that with
simultaneous contributions from gluino and chargino one may be able
to explain these discrepancies.

In section 6 we focus on the CP asymmetries in $B\to K \pi$
processes. We show that with a large gluino contribution it is quite
natural to account for the recent experimental results of direct CP
asymmetries. The SUSY contributions to the mixing CP asymmetry of
$B^0 \to K^0 \pi^0$ is also discussed. Finally, section 7 contains
our main conclusions.
%
\section{$B\to K \pi$ in the Standard Model}
In this section we analyze the SM predictions for the CP asymmetries
and the branching ratios of $B\to K \pi$ decays. The effective
Hamiltonian of $\Delta B=1$ transition governing these processes can
be expressed as
\begin{eqnarray}
H^{\Delta B=1}_{\rm eff}&\!=\!&\frac{G_F}{\sqrt{2}} \sum_{p=u,c}
\lambda_p \Big(C_1 Q_1^p + C_2 Q_2^p + \sum_{i=3}^{10}C_i Q_i +
C_{7\gamma} Q_{7\gamma} + C_{8g} Q_{8g}\Big)  +h.c.,~~
\label{Heff}
\end{eqnarray}
where $\lambda_p= V_{pb} V^*_{ps}$ and $C_i$ are the Wilson
coefficients and $Q_i$ are the relevant local operators which can
be found in Ref.\cite{bbl}. Within the SM, the $b\to s$ transition
can be generated through exchange of $W$-boson. The Wilson
coefficients which describes such a transition can be found in
Ref.\cite{bbl}

The calculation of the decay amplitudes of $B\to K \pi$ involves
the evaluation of the hadronic matrix elements of the above
operators in the effective Hamiltonian, which is the most
uncertain part of this calculation. Adopting the QCD factorization
\cite{BN}, the matrix elements of the effective weak Hamiltonian
can be written as \be \langle \pi K\vert H_{eff} \vert \bar{B}
\rangle = \frac{G_F}{\sqrt{2}}\sum_{p=u,c} \lambda_p \langle \pi K
\vert \left(\mathcal{T}_p + \mathcal{T}_p^{ann}\right) \vert
\bar{B} \rangle , \ee where \be \langle \pi K \vert \mathcal{T}_p
\vert \bar{B} \rangle = \sum_{i=1}^{10} a_i(\pi K) \langle \pi K
\vert Q_i \vert \bar{B} \rangle_F , \ee and \be \langle \pi K
\vert \mathcal{T}_p^{ann} \vert \bar{B} \rangle = f_B f_K f_{\pi}
\sum_{i=1}^{10} b_i(\pi K) .\ee The term $\mathcal{T}_P$ arises
from the vertex corrections, penguin corrections and hard
spectator scattering contributions which are involved in the
parameters $a_i(\pi K)$. The $\langle \pi K \vert Q_i \vert
\bar{B} \rangle_F$ are the factorizable matrix elements, i.e. if
any operator $Q= j_1 \bigotimes j_2$, then $\langle \pi K \vert
Q_i \vert \bar{B} \rangle_F = \langle \pi \vert j_1 \vert \bar{B}
\rangle \langle K \vert j_2 \vert 0\rangle$ or $\langle K \vert
j_1 \vert \bar{B} \rangle \langle \pi \vert j_2 \vert 0\rangle$.
The other term $\mathcal{T}^{ann}_p$ includes the weak
annihilation contributions which are absorbed in the parameters
$b_i(\pi K)$. Following the notation of Ref.\cite{BN} we write the
decay amplitude of $B\to K \pi$ as: \bea
A_{B^-\to\pmkz}&=&\sum_{p=u,c}\lambda_pA_{\pi
\Kbar}\left[\delta_{pu}\beta_2+\alpha_4^p-
\frac{1}{2}\alpha^p_{4,EW}+\beta_3^p+\beta^p_{3,EW}\right] \label{ampl1}\\
\sqrt{2}A_{B^-\to\pzkm}&=&\sum_{p=u,c}\lambda_pA_{\pi\Kbar}\left[\delta_{pu}(\alpha_1+\beta_2)+
\alpha_4^p+\alpha^p_{4,EW}+\beta^p_3+\beta^p_{3,EW}\right] \nonumber\\
&+&\sum_{p=u,c}\lambda_pA_{\Kbar\pi}\left[\delta_{pu}\alpha_2+\frac{3}{2}\alpha^p_{3,EW}
\right] \label{ampl2}\\
A_{\overline{B}^0\to\ppkm}&=&\sum_{p=u,c}\lambda_pA_{\pi\Kbar}\left[\delta_{pu}\alpha_1+
\alpha^p_4+\alpha^p_{4,EW}+\beta^p_3-\frac{1}{2}\beta^p_{3,EW}\right]\label{ampl3}\\
\sqrt{2}A_{\overline{B}^0\to
\pzkz}&=&\sum_{p=u,c}\lambda_pA_{\pi\Kbar}\left[-\alpha^p_4+
\frac{1}{2}\alpha^p_{4,EW}-\beta^p_3+\frac{1}{2}\beta^p_{3,EW}\right] \nonumber\\
&+&\sum_{p=u,c}\lambda_pA_{\Kbar\pi}\left[\delta_{pu}\alpha_2+\frac{3}{2}\alpha^p_{3,EW}\right].
\label{ampl4} \eea Here the coefficients of the flavor operators
$\alpha_i^p(\pi K)$ and $\beta_i^p(\pi K)$ are given in terms of
the coefficients $a_i^p(\pi K)$ and $b_i^p(\pi K)$ respectively
\cite{BN}. The parameter $A_{\pi \bar{K}}$ ($A_{\bar{K}\pi}$) is
given by $i \frac{G_F}{\sqrt{2}} m_B^2 F_0^{B\to \pi (K)}
f_{K(\pi)}$. Note that the parameters $b_i$ of the weak
annihilation and hard scattering contributions contain infrared
divergence which are usually parameterized as \be X_{A,H} \equiv
\left(1+ \rho_{A,H} e^{i\phi_{A,H}}\right)
\ln\left(\frac{m_B}{\Lambda_{h}}\right), \ee where $\rho_{A,H}$
are free parameters to be of order one, $\phi_{A,H} \in[0,2\pi]$,
and $\Lambda_h = 0.5$. As discussed in
Ref.\cite{Gabrielli:2004yi}, the experimental measurements of the
branching ratios impose upper bound on the parameter $\rho_A$. If
one does not assume fine tuning between the parameters $\rho$ and
$\phi$, the typical upper bound on $\rho_A$ is of order of $\rho_A
\lsim 2$.

Fixing the experimental and the SM parameters to their center
values, one can determine the explicit dependence of the decay
amplitudes of the $B \to K \pi$ on the corresponding Wilson
coefficients. For instance, with $\gamma=\pi/3$, and $\rho_{A,H}$
and $\phi_{A,H}$ are of order one, the decay amplitude of
$\bar{B}^0\to K^- \pi^+ $ is given by \bea A_{\bar{B}^0\to \pi^+K^-}
\times 10^{8} &\simeq& (1.05 - 0.02~i)C_1 + (0.24 + 0.07~i)C_2 +
(3.1+14.5~i)C_3\no &+&(4.9+37.7~i)C_4 - (2.9 -13.1~i)C_5 + (5.5 -
43.7~i)C_6 \no
&+&(1.7+10.4~i) C_7 + (5.8+36.5~i) C_8 + (2.8 + 12.7~i)C_9\label{num+-}\\
&+& (0.6+ 35.5~i) C_{10} - (0.0006 + 0.04~i) C_{7\gamma}^{eff} -
(0.04 + 2.5~i) C_{8g}^{eff}\nonumber.\eea Similar expression can be
obtained for  $\bar{B}^0\to K^0 \pi^0 $: \bea A_{\bar{B}^0\to \pi^0
K^0} \times 10^{8} &\simeq& (-0.14 + 0.3~i)C_1 + (0.4 + 0.2~i)C_2 -
(2.2 + 10.5~i)C_3\no &-&(3.5 + 26.7~i)C_4 + (2.1 -9.3~i)C_5 + (3.9 -
30.9~i)C_6 \no
&-&(1.7 + 37.02~i) C_7 - (1.9 + 1.7~i) C_8 + (1.3 + 46.6~i)C_9\\
&+& (2.2 + 28.8~i) C_{10} - (0.0002 + 0.01~i) C_{7\gamma}^{eff} +
(0.03 + 1.8~i) C_{8g}^{eff}.\nonumber \label{num00}\eea The
amplitude of $B^-\to K^- \pi^0 $ can be written as \bea A_{B^-\to
\pi^0 K^-} \times 10^{8} &\simeq& (0.9 + 0.06~i)C_1 + (0.6 +
0.3~i)C_2 + (2.2 + 10.5~i)C_3\no &+&(3.5 + 26.7~i)C_4 - (2.1
-9.3~i)C_5 - (3.9 - 30.9~i)C_6 \no
&-&(2.7 + 32.4~i) C_7 - (5.4 - 17.8~i) C_8 + (1.9 + 51.6~i)C_9\\
&+& (2.4 + 42.8~i) C_{10} - (0.0004 + 0.03~i) C_{7\gamma}^{eff} -
(0.03 + 1.8~i) C_{8g}^{eff}.\nonumber \label{num0-}\eea Finally, the
amplitude of $B^-\to K^0 \pi^- $ is given by \bea A_{B^-\to \pi^-
K^0}\times 10^{8} &\simeq& (0.4 - 0.4~i)C_1 -(0.00004 - 0.02~i)C_2 +
(3.1 + 14.9~i)C_3\no &+&(4.9+37.7~i)C_4 - (2.9 -13.1~i)C_5 + (5.5 -
43.7~i)C_6 \no
&-&(3.2 + 3.8~i) C_7 - (10.8 + 13.8~i) C_8 - (1.9 + 5.7~i)C_9\label{num-0}\\
&-& (0.3+ 19.7~i) C_{10} + (0.0003 + 0.02~i) C_{7\gamma}^{eff} -
(0.04 + 2.5~i) C_{8g}^{eff},\nonumber \eea where $C_{7\gamma}^{eff}=
C_{7\gamma} -\frac{1}{3} C_5 - C_6$ and $C_{8g}^{eff} = C_{8g} +
C_5$. The SM contributions to the Wilson coefficients of $b\to s$
transition, which are the relevant ones for $B\to K \pi$, are given
by \bea {\bf C}_1^{SM} &\simeq& 1.077, ~~~~ {\bf C}_2^{SM} \simeq
-0.175, ~~ {\bf C}_3^{SM} \simeq 0.012, ~~ {\bf C}_4^{SM} \simeq
-0.33, ~~ {\bf C}_5^{SM} \simeq 0.0095, \no {\bf C}_6^{SM} &\simeq&-
0.039, ~~ {\bf C}_7^{SM} \simeq 0.0001, ~~ {\bf C}_8^{SM} \simeq
0.0004, ~~ {\bf C}_9^{SM} \simeq -0.01,~~ {\bf C}_{10}^{SM} \simeq
0.0019, \no {\bf C}_{7\gamma}^{SM} &\simeq& -0.315, ~~ {\bf
C}_{8g}^{SM} \simeq -0.149. \label{smC}\eea

From these values, it is clear that within the SM, the dominant
contribution to the $B\to K \pi$ decay amplitudes comes from the QCD
penguin operator $Q_4$. However the QCD penguin preserves the
isospin. Therefore, this contribution is the same for all the decay
modes. Isospin violating contributions to the decay amplitudes arise
from the current-current operators $Q^u_1$ and $Q^u_2$ which are
called 'tree' contribution and from the electroweak penguins which
are suppressed by a power $\alpha/\alpha_s$. As can be seen from the
coefficients of $C_{7-10}$ in Eqs.(\ref{num+-}-\ref{num-0}), the
electroweak penguin contributions to the amplitudes of $B\to K \pi$
could be in general sizable and non universal. However, due to the
small values of the corresponding Wilson coefficients in the SM
(\ref{smC}), these contributions are quite suppressed.

Note also that the $Q_1$ contribution to $A_{\bar{B}^0\to \pi^+
K^-}$ and $A_{B^-\to \pi^0 K^-}$ is one order of magnitude larger
than its contribution to the other two decay amplitudes. Therefore,
in the SM, the amplitudes $A_{\bar{B}^0\to \pi^+ K^-}$ and
$A_{B^-\to \pi^0 K^-}$ can be approximated as function of $C_1$ and
$C_4$, while the amplitudes $A_{\bar{B}^0\to \pi^0 K^0}$ and
$A_{B^-\to \pi^- K^0}$ are approximately given in terms of $C_4$
only. It is worth noting that the difference between the
coefficients of $C_4$ in the amplitudes $A_{\bar{B}^0\to \pi^+ K^-}$
and $A_{B^-\to \pi^0 K^-}$ is just due to the factor of $\sqrt{2}$
in Eq.(\ref{ampl2}), which is the same difference between the
corresponding coefficients in $A_{\bar{B}^-\to \pi^- K^0}$ and $-
A_{B^0\to \pi^0 K^0}$.

We are now in a position to determine the SM results for the CP
asymmetries  and the CP average branching ratios of $B\to K \pi$
decays within the framework of the QCD factorization
approximation. The direct CP violation may arise in the decay
$B\to K \pi$ from the interference between the tree and penguin
diagrams. The direct CP asymmetry of $B^0\to K^- \pi^+$ decay
$A^{CP}_{K^-\pi^+}$ is defined as \be A^{CP}_{K^-\pi^+} =
\frac{\left\vert A(B^0 \to K^- \pi^+)\right\vert^2 - \left\vert
A(\bar{B}^0 \to K^+ \pi^-)\right\vert^2}{ \left\vert A(B^0 \to K^-
\pi^+)\right\vert^2 + \left\vert A(\bar{B}^0 \to K^+
\pi^-)\right\vert^2},\ee and similar expressions for the
asymmetries $A^{CP}_{\bar{K}^0\pi^-}$, $A^{CP}_{K^-\pi^0}$ and
$A^{CP}_{\bar{K}^0\pi^0}$. Also the branching ratio can be written
in terms of the corresponding decay amplitude as \be BR(B\to K
\pi) = \frac{1}{8 \pi} \frac{\vert P\vert}{M_B^2} \vert A(B\to K
\pi) \vert^2 \frac{1}{\Gamma_{tot}}, \ee where \be \vert P \vert =
\frac{\left[\left(M_B^2 - (m_K + m_{\pi})^2\right)\left(M_B^2 -
(m_K - m_{\pi})^2\right)\right]^2}{2 M_B} \ee

The SM results are summarized in Tables 2 and 3. In Table 2, we
present the predictions for the branching ratios of the four decay
modes of $B\to \pi K$. We assume that $\gamma =\pi/3$ and consider
some representative values of $\rho_{A,H}$ and $\phi_{A,H}$ to check
the corresponding uncertainty. Namely, $\rho_{A,H} = 0,1,3$ and
$\phi_{A,H}= \mathcal{O}(1)$ are considered.
\begin{table}[t]
\centering
\begin{tabular}{|c|c|c|c|}
\hline
Branching ratio &$\rho_{A,H}=0$ & $\rho_{A,H}=1$ \&
$\phi_{A,H}\sim 1$ & $\rho_{A,H}=3$ \& $\phi_{A,H}\sim 1$\\
\hline
$BR_{\bar{K}^0\pi^-}\times 10^6$ & $31.06$ & $33.35$ & $43.92$ \\
\hline
$BR_{K^- \pi^0}\times 10^6$ & 17.31 & 18.45 & 23.36 \\
\hline
$BR_{K^-\pi^+}\times 10^6$ & 25.87 & 27.98 & 39.55 \\
\hline
$BR_{\bar{K}^0\pi^0}\times 10^6$  & 11.41 & 12.47 & 18.66\\
\hline
$R_n$ & 1.13 & 1.12 & 1.059 \\
\hline
$R_c$ & 1.11 & 1.106 & 1.063 \\
\hline
$R_n$ & 0.83 & 0.838 & 0.9 \\
\hline
\end{tabular}
\caption{The SM predictions for the branching ratios of the four
decay modes of $B\to K\pi$ with $\gamma =\pi/3$} \label{table2}
\end{table}
From these results, one can see that for $\rho_{A,H} \in [0,1]$
the SM predicted values for the branching ratios of $B\to K \pi$
are less sensitive to the hadronic parameters. Larger values of
$\rho_{A,H}$ enhance the branching ratios and eventually they
exceed the experimental limits presented in table 1 for $\rho
> 2$. It is also remarkable that the SM results for the $BR(B^-
\to \bar{K}^0\pi^- )$ and $BR(\bar{B}^0 \to \bar{K}^0\pi^0 )$ are
larger than the experiment measurements, while the results for
$BR(B^- \to K^-\pi^0 )$ and $BR(\bar{B}^0 \to K^-\pi^+ )$ are
consistent with their experimental values. This discrepancy does
not seem to be resolved in the SM, even if we consider large
hadronic uncertainties. The parameters $R_c$ and $R_n$, defined in
Eqs.(\ref{Rcresult},\ref{Rnresult}) as the ratio of the CP average
branching ratios of $B\to K \pi$ exhibit this deviation from the
SM prediction in a clear way. The results in Table 2 show that in
the SM $R_c \simeq R_n > 1$. However, the recent experimental
measurements reported in Table 1, implies that $R_c \sim 1$ and
$R_n < 1$. It is very difficult to have this situation within the
SM. As emphasized above, in the SM the amplitudes of $B\to K \pi$
can be approximately written as \bea A_{\bar{B}^0\to \pi^+ K^-}
&\simeq&
(a_1 + b_1~i)~C_1 + (a_2+b_2~i)~C_4 , \\
A_{\bar{B}^0\to \pi^0 K^0} &\simeq& - \frac{1}{\sqrt{2}}(a_2 +
b_2~i)~C_4 ,\\
A_{B^-\to \pi^0 K^-} &\simeq& \frac{1}{\sqrt{2}} (a_1 + b_1~i)~C_1
+ \frac{1}{\sqrt{2}}(a_2 + b_2~i)~C_4 \\
A_{B^-\to \pi^- K^0} &\simeq& (a_2+b_2~i)~C_4.\eea Thus, the
parameters $R_c$ and $R_n$ are given by \be R_c = R_n =
\frac{\left\vert (a_1 + b_1~ i)~C_1 + (a_2 + b_2~ i)~C_4
\right\vert^2}{\left\vert (a_2 + b_2~ i)~C_4 \right\vert^2} ~
\gsim 1 \ee which is consistent with the result given in Table 2,
using the full set of the Wilson coefficients.

\begin{table}[t]
\centering
\begin{tabular}{|c|c|c|c|}
\hline CP asymmetry & $\rho_{A,H}=0$ & $\rho_{A,H}=1$ \&
$\phi_{A,H}\sim 1 (-1)$ & $\rho_{A,H}=3$ \& $\phi_{A,H}\sim 1(-1)$\\
\hline
$A^{CP}_{\bar{K}^0\pi^- }$ & 0.007 & 0.0086 (0.005) & 0.0078 (0.001) \\
\hline
$A^{CP}_{K^-\pi^0 }$ & 0.029 & 0.063 (-0.006) & 0.185 (-0.15)\\
\hline
$A^{CP}_{ K^-\pi^+}$ & 0.0044 & 0.057 (-0.049) & 0.194 (-0.19) \\
\hline
$A^{CP}_{\bar{K}^0\pi^0 }$  & -0.02 & -0.013 (-0.025)  & -0.019 (-0.002)\\
\hline
\end{tabular}
\caption{The SM predictions for the direct CP asymmetries of the
four decay modes of $B\to \pi K$ with $\gamma =\pi/3$}
\label{table2}
\end{table}

Now we turn to the SM predictions for the CP asymmetries of $B\to
K \pi$. Let us start by considering the approximation that the
decay amplitudes for $B^- \to K^0\pi^- $ and $\bar{B}^0 \to
K^0\pi^0 $ are dominated by the pure gluon penguin operator $Q_4$
while the amplitudes for $B^- \to K^-\pi^0 $ and $\bar{B}^0 \to
K^- \pi^+$ are given by $Q_4$ and also by the tree contribution of
the current-current operator $Q_1$. In this case, the following
results are expected: The direct CP asymmetries $A^{CP}_{K^0 \pi^-
}$ and $A^{CP}_{K^0\pi^0}$ should be very tiny (equal zero in the
exact limit of this approximation). The direct CP asymmetries
$A^{CP}_{K^0\pi^0}$ and $A^{CP}_{ K^-\pi^+}$ should be of the same
order and larger than the other two asymmetries.

The SM results of the CP asymmetries for the different decay
modes, including the effect of all local operators $Q_i$, are
given in Table 3. As in the case of the branching ratios, we
assume that $\gamma =\pi/3$, and $\rho_{A,H} =0,1,3$. Respect to
the strong phases $\phi_{A,H}$, we take it to be of order one as
before. Due to the sensitivity of the CP asymmetry on their sign,
we consider both cases of $\phi_{A,H}= \mathcal{O}(\pm 1)$. Few
comments on the results of the direct CP asymmetries given in
Table 2 are in order:
\begin{enumerate}
\item The CP asymmetries $A^{CP}_{K^-\pi^0 }$ and $A^{CP}_{K^-\pi^+}$
are sensitive to the sign $\phi_A$ (note that $\phi_H$ is
irrelevant for these processes). On the contrary, the CP
asymmetries $A^{CP}_{K^0\pi^-}$ and $A^{CP}_{K^0\pi^0}$ are
insensitive to this sign.

\item  As expected, the results of the CP asymmetries
$A^{CP}_{K^0\pi^-}$ and $A^{CP}_{K^0\pi^0}$ are very small even
with large values of $\rho_A$.

\item  The value of $A^{CP}_{K^-\pi^0}$ and $A^{CP}_{K^-\pi^+}$ can be
enhanced by considering large value of $\rho_A$ and one gets values
for $A^{CP}_{K^-\pi^+}$ of order the experimental result given in
Table 1. However, it is very important to note that in this case,
the CP asymmetry $A^{CP}_{K^-\pi^0}$ is also enhanced in the same
way and it becomes one order of magnitude larger than its
experimental value. \end{enumerate}

While a confirmation with more accurate experimental data is
necessary, the above results of the branching ratio and the direct
CP asymmetries of $B\to \pi K$ show that within the SM the current
experimental measurements listed in Table 1 do not seems to be
accommodated even if one considers large hadronic uncertainties.
It is worth stressing that the QCD correction would not play an
essential role in solving this $K \pi$ puzzles. Furthermore, since
we are interested here in the ratio of the amplitudes, many of the
theoretical uncertainties cancel. So it can not be the source of
these discrepancies.

Another useful way of parameterizing the decay amplitudes can be
obtained by factorizing the dominant penguin amplitude $P$, where
$P$ is defined as \cite{BBNS2} \be P
e^{i\delta_P}=\alpha_4^c-\frac{1}{2}\alpha^c_{4,EW}+\beta_3^c+\beta^c_{3,EW}.
\label{eq:P-def} \ee In this case, one can write the above
expressions for the decay amplitude as follows: \bea
A_{B^-\to\pmkz}&=&\lambda_cA_{\pi\Kbar}P\left[1 +r_A
e^{i\delta_A}e^{-i\gamma}\right], \no
\sqrt{2}A_{B^-\to\pzkm}&=&\lambda_cA_{\pi\Kbar}P\left[1 +\left(r_A
e^{i\delta_A} + r_C e^{i \delta} \right) e^{-i\gamma} + r_{EW}
e^{i\delta_{EW}}\right], \no A_{\overline{B}^0\to\ppkm}&=&
\lambda_cA_{\pi\Kbar}P\left[1+\left(r_A e^{i\delta_A}+ r_T
e^{i\delta_T}\right)e^{-i\gamma} + r_{EW}^C e^{i\delta_{EW}^C}\right], \label{paramerization1}\\
-\sqrt{2}A_{\overline{B}^0\to\pzkz}&\!=\!&\lambda_cA_{\pi\Kbar}P\left[1
\!+\! \left( r_A e^{i\delta_A} + r_T e^{i\delta_T} - r_C
e^{i\delta_C}\right) e^{-i\gamma} + r_{EW}^C e^{i\delta_{EW}^C} -
r_{EW} e^{i\delta_{EW}}\right], \nonumber \eea where \bea r_A
e^{i\delta_A}&=&\epsilon_{KM}\left[\beta_2+\alpha_4^u-\frac{1}{2}\alpha^u_{4,EW}+
\beta_3^u+\beta^u_{3,EW}\right]/P,\\
r_{T}e^{i\delta_T}&=&\epsilon_{KM}\left[
\alpha_1+\frac{3}{2}\alpha^u_{4,EW}-
\frac{3}{2}\beta^u_{3,EW}-\beta_2) \right]/P, \\
r_C e^{i\delta_P}&=&\epsilon_{KM}\left[ \alpha_1 +
R_{K\pi}\alpha_2+\frac{3}{2}(R_{K\pi}
\alpha^u_{3,EW}+\alpha^u_{4,EW}) \right]/P ,\\
r_{EW} e^{i\delta_{EW}}&=&\left[\frac{3}{2}(R_{K\pi}\alpha^c_{3,EW}+\alpha^c_{4,EW})
\right]/P, \\
r^C_{EW} e^{i\delta_{EW}^C}&=&
\left[\frac{3}{2}(\alpha^c_{4,EW}-\beta^c_{3,EW})\right]/P. \eea
Here we define
$\lambda_u/\lambda_c\equiv\epsilon_{KM}e^{-i\gamma}$,
$R_{K\pi}=A_{\pi\Kbar}/A_{\Kbar \pi}$, and $\delta_A$, $\delta_T$,
$\delta_C$, $\delta_{EW}$ and $\delta_{EW}^C$ as strong
interaction phases. The SM contributions within the QCD
facorization leads to the following results:
 \bea
(Pe^{i\delta_P})_{\sm}&=& -0.11 e^{0.051 i},~~~~~~~~~ (r_A
e^{i\delta_A})_{\sm}= 0.019 e^{0.26 i},\no
(r_{C}e^{i\delta_{C}})_{\sm}&=& 0.186 e^{2.9 i},~~~~~~~ (r_T
e^{i\delta_T})_{\sm}=0.191 e^{2.9 i },\no (r_{EW} e^{i
\delta_{EW}})_{\sm}&=& 0.13 e^{- 0.2 i},~~~~~~~~~~~~ (r_{EW}^C
e^{i\delta_{EW}^C})_{\sm}= 0.012 e^{- 2.5 i }~. \label{eq:qCSM}
\eea

As can be seen from this result, within the SM $r_A$ and
$r_{EW}^C$ are much smaller than $r_C$, $r_T$ and $r_{EW}$, so
that they can be easily neglected. In this case, the parameters
$R_c$ and $R_n$ can be expressed by the following approximated
expressions \bea R_c & \simeq & 1 + 2 r_C \cos\delta_C \cos\gamma + 2 r_{EW} \cos\delta_{EW}, \\
R_n &\simeq& \frac{1 + 2 r_T \cos \delta_T \cos\gamma}{1 + 2 r_T
\cos \delta_T \cos \gamma - 2 r_C \cos \delta_C \cos \gamma - 2
r_{EW} \cos \delta_{EW}}, \eea which confirms our previous
conclusion that in the SM $R_n \sim R_c \gsim 1$. Explicitly,
using the results of Eq.(\ref{eq:qCSM}), one finds that \be
R_c=1.08(1.45), \hspace{1cm}R_n=1.13(1.6),
\hspace{1cm}R=0.757(0.673) \ee for $\gamma=\pi /3 (2\pi /3)$,
which is quite close to the full result that we obtained in Table
2, with $\rho_A \sim 1$.

Now, we would like to comment on the mixing CP asymmetry of $B\to K
\pi$. CP violation in the interference between mixing and decay can
be observed as time dependent oscillation of the CP asymmetry. The
amplitude of the oscillation in charmonium decay modes provides a
theoretical clean determination of the parameter $\sin 2 \beta$ of
the unitary triangle. The SM predicts the $B$-decay modes, dominated
by a single penguin amplitude such that $B\to \phi K$, $B\to \eta'
K$ and $B \to K^0 \pi^0$ to have the same time dependent CP
asymmetry equal to $\sin 2 \beta$. Again this result contradicts the
experimental measurement given in Table 1. Note that the latest
experimental results on the mixing CP asymmetry of $B\to \phi K_S$
process are given by \cite{giorgi,sakai} \bea S_{\phi K_S}&=&0.50\pm
0.25^{+0.07}_{-0.04}\;\;
({\rm BaBar}),\nonumber\\
&=&0.06\pm 0.33 \pm 0.09\;\; ({\rm Belle})\, , \label{Sphi} \eea
where the first errors are statistical and the second systematic.
Thus, the average of this CP asymmetry is $S_{\phi K_S}=0.34 \pm
0.20$. On the other hand, the most recent measured CP asymmetry in
the $B^0\to \eta^{\prime} K_S$ decay is found by BaBar
\cite{giorgi} and Belle \cite{sakai} collaborations as \bea
S_{\eta^{\prime} K_S}&=&0.27\pm 0.14\pm 0.03 \;\; ({\rm BaBar}) \nonumber\\
&=&0.65\pm 0.18\pm 0.04 \;\; ({\rm Belle}), \label{Seta} \eea with
an average $S_{\eta^{\prime} K_S}= 0.41 \pm 0.11$, which shows a
2.5$\sigma$  discrepancy from the SM expectation. This difference
among $S_{\phi K}$, $S_{\eta' K}$, $S_{K^0\pi^0}$ and $\sin 2
\beta$ is also considered as a hint for new physics beyond the SM,
in particular for supersymmetry.

\section{$B\to K \pi$ in SUSY models}

As mentioned in the previous section, due to the asymptotic
freedom of QCD, the calculation of the hadronic decay amplitude of
$B\to K \pi$ can be factorized by the product of long and short
distance contributions. The short distance contributions,
including the SUSY effects are contained in the Wilson
coefficients $C_i$.

The SUSY contributions to the $b\to s$ transition could be dominated
by the gluino or the chargino intermediated penguin diagrams
\cite{Gabrielli:2004yi}. It turns out that the dominant effect in
both contributions is given by chromomagnetic penguin ($Q_{8g}$).
However in case of $B\to K \pi$, it was observed that this process
is more sensitive to the isospin violating interactions
\cite{Khalil:2004yb,kpi-puzzle}, namely the contributions from the
electromagnetic penguin ($Q_{7\gamma}$) and photon- and $Z$-penguins
contributions to $Q_7$ and $Q_9$. Therefore, in our discussion we
will focus only on these contributions, although in our numerical
analysis we keep all the contributions of the gluino and chargino.

For the gluino exchange, it turns out that the $Z$-penguin
contributions to $C_{7,9}$ are quite small and can be neglected
with respect to the photon-penguin contributions. At the first
order in the mass insertion approximation, the gluino
contributions to the Wilson coefficients $C_{7\gamma,8g}$, $C_7$
and $C_{9}$ at SUSY scale $M_S$ are given by
\begin{eqnarray}
C^{\tilde{g}}_7 (M_S)&=&C_9(M_S) = \frac{2 \alpha_s \alpha}{9
\sqrt{2} G_F m_{\tilde{q}}^2}~
\frac{1}{3}(\delta_{LL}^d)_{23} P_{042}(x,x),\label{C7}\\
C^{\tilde{g}}_{7\gamma}(M_S)&=&\frac{8 \alpha_s \pi}{9 \sqrt{2}
G_F m_{\tilde{q}}^2} \left[ (\delta_{LL}^d)_{23} M_3(x) +
(\delta_{LR}^d)_{23} \frac{m_{\tilde{g}}}{m_b}
M_1(x)\right] ,\label{C7g}\\
C^{\tilde{g}}_{8g}(M_S)&=&\frac{\alpha_s \pi}{\sqrt{2} G_F
m_{\tilde{q}}^2}\Big[ (\delta_{LL}^d)_{23}\left( \frac{1}{3}
M_3(x) + 3 M_4(x)\right) \no &+&
(\delta_{LR}^d)_{23}\frac{m_{\tilde{g}}}{m_b} \left(\frac{1}{3}
M_3(x) + 3 M_2(x)\right)\Big] ,\label{C8g}
\end{eqnarray}
where $x=m_{\tilde{g}}^2/m_q^2$ and the functions $M_1(x),M_2(x)$
and $P_{ijk}(x,x)$ can be found in Ref.\cite{ggms,silvestrini}.
The coefficients $\tilde{C}_{7\gamma,8g}$ and $\tilde{C}_{7,9}$
are obtained from $C_{7\gamma,8g}$ and $C_{7,9}$ respectively, by
the chirality exchange $L \leftrightarrow R$. As can be seen from
Eqs.(\ref{C7g},\ref{C8g}), the term proportional to
$(\delta^d_{LR})_{23}$ in the coefficients $C_{7\gamma,8g}$ has a
large enhancement factor $m_{\tilde{g}}/m_b$. This enhancement
factor is responsible for the dominant gluino effects in
$B$-decays, although this mass insertion is strongly constrained
from $b\to s \gamma$. Note also that, since the photon-penguin
gives the same contributions to $C_7$ and $C_9$, and we neglect
the $Z$-penguin contributions, we have $C_7=C_9$. Finally, it is
clear that the coefficients $C_{7,9}$ is suppressed with respect
to $C_{7\gamma,8g}$ by a factor $\alpha/4\pi$ at least.

It is worth mentioning that the mass insertion
$(\delta_{LR}^d)_{23}$ can be generated by the mass insertion
$(\delta_{LL}^d)_{23}$ as follows \cite{Abel:2004te}$$
(\delta_{LR}^d)_{23}= (\delta_{LL}^d)_{23} ~
(\delta_{LR}^d)_{33},$$ where
$$(\delta_{LR}^d)_{33} \sim \frac{m_b (A_b -\mu \tan
\beta)}{m_{\tilde{d}}^2} \sim \frac{m_b}{m_{\tilde{d}}^2}~ \tan
\beta \sim 10^{-2} \tan \beta .$$ Therefore,
$$(\delta_{LR}^d)_{23} \simeq 10^{-2} \tan \beta
(\delta_{LL}^d)_{23}.$$ Hence, for a moderate value of $\tan
\beta$ and $(\delta_{LL}^d)_{23} \sim \mathcal{O}(0.1)$, one
obtains $(\delta_{LR}^d)_{23}$ of order $10^{-2}$, which can
easily imply significant contributions for the $S_{\phi K}$ and
also account for the different results between $S_{\phi K}$ and
$S_{\eta' K}$. Thus in our analysis we define \be
(\delta_{LR}^d)_{23_{\mathrm{eff}}} = (\delta_{LR}^d)_{23} +
(\delta_{LL}^d)_{23}~ (\delta_{LR}^d)_{33}. \ee

It is important to stress that in case of
$(\delta_{LR}^d)_{23_{\mathrm{eff}}}$ dominated by double mass
insertions, we still call this scenario as $LR$ contribution. This
is due to the fact that the main SUSY contribution is still
through the $C_{8g}$ which is enhanced by the chirality flipped
factor $m_{\tilde{g}}/m_b$. In the literatures \cite{yamaguchi},
this contribution has been considered in analyzing the CP
asymmetry of $B\to \phi K$ and it was called as LL contribution,
as indication for the large mixing in the squark mass matrix and
dominant effect of $(\delta_{LL}^d)_{23}$. However, we prefer to
work with the notation $LR_{\mathrm{eff}}$ to be able to trace the
effective operators that may lead to dominant contributions for
different $B$ decay channels.

The dominant chargino contributions are found to be also due to
the chromomagnetic penguin, magnetic penguin and $Z$-penguin
diagrams. As emphasized in Ref.\cite{Gabrielli:2004yi}, these
contributions depend on the up sector mass insertion
$(\delta^u_{LL})_{32}$ and $(\delta^u_{RL})_{32}$ while the  $LR$
and $RR$ contributions are suppressed by $\lambda^2$ or
$\lambda^3$, where $\lambda$ is the Cabibbo mixing. At the first
order in the mass insertion approximation, the chargino
contributions to the Wilson coefficients are given by
\cite{Gabrielli:2004yi} \bea C^{\chi}_7 (M_S)&=&
\frac{\alpha}{6\pi} \left(4 C_{\chi} +
D_{\chi}\right),\\
C^{\chi}_9 (M_S)&=& \frac{\alpha}{6\pi} \left(4
(1-\frac{1}{\sin^2\theta_W}) C_{\chi} +
D_{\chi}\right),\\
C^{\chi}_{7\gamma} &=& M_{\gamma},\\
C^{\chi}_{8g} &=& M_{g}, \eea where the functions $F\equiv
C_{\chi}$ ($Z$-penguin), $D_{\chi}$ (photon-penguin), $M^{\gamma}$
(magnetic-penguin), and $M^{g}$ (chromomagnetic penguin) are given
by \cite{Gabrielli:2004yi}
\begin{eqnarray}
F_{\chi} =  \Big[\;\du{LL}{32} + \lambda \du{LL}{31}\;\Big] R_F^{LL}
+ \Big[\;\du{RL}{32} + \lambda \du{RL}{31}\;\Big]\, Y_t\, R_F^{RL}.
\label{Fchapprox}
\end{eqnarray}
The functions $R^{LL}_F$ and $R^{RL}_F$, $F$ depend on the SUSY
parameters through the chargino masses ($m_{\chi_i}$), squark
masses ($\tilde{m}$)and the entries of the chargino mass matrix.
For the $Z$ and magnetic (chromomagnetic) dipole penguins
$R_C^{LL,RL}$ and $R_{M^{\gamma (g)}}^{LL,RL}$ respectively, we
have \cite{Gabrielli:2004yi}
\begin{eqnarray}
R_C^{LL}&=& \sum_{i=1,2}|V_{i1}|^2 \, P_C^{(0)}(\bar x_i)
+\sum_{i,j=1,2} \left[ U_{i1}V_{i1}U_{j1}^{\star}V_{j1}^{\star}\,
P_C^{(2)}(x_i,x_j) \right.
\nonumber\\
&&+ \left. |V_{i1}|^2 |V_{j1}|^2 \left(\frac{1}{8}-P_C^{(1)}(x_i,
x_j)\right)\right], \nonumber
\\
R_C^{RL}&=&-\frac{1}{2} \sum_{i=1,2}\, V_{i2}^{\star}V_{i1}\,
P_C^{(0)}(\bar x_i,\bar{x}_{it}) - \sum_{i,j=1,2}\,
V_{j2}^{\star}V_{i1}\left( U_{i1}U_{j1}^{\star}\,
P_C^{(2)}(x_i,x_{it}, x_j,x_{jt})
\right.\nonumber \\
&&+ \left. V_{i1}^{\star} V_{j1}\, P_C^{(1)}(x_i, x_j)\right),
\nonumber
\\
R_{M^{\gamma, g}}^{LL}&=&\sum_i |V_{i1}|^2\, x_{Wi}\,
P_{M^{\gamma,g}}^{LL}(x_i) - Y_b\sum_i V_{i1} U_{i2}\, x_{Wi}\,
\frac{m_{\chi_i}}{m_b} P_{M_{\gamma,g}}^{LR}(x_i), \nonumber
\\
R_{M^{\gamma, g}}^{RL}&=& -\sum_i V_{i1}V_{i2}^{\star}\, x_{Wi}\,
P_{M_{\gamma,g}}^{LL}(x_i,x_{it}), \label{Rterms}
\end{eqnarray}
where $Y_b$ is the Yukawa coupling of bottom quark, $x_{\W
i}=m_W^2/m_{\chi_i}^2$, $x_{i}=m_{\chi_i}^2/\tilde{m}^2$, $\bar
x_i =\tilde{m}^2/m_{\chi_i}^2$, and
$x_{it}=m_{\chi_i}^2/{m^2_{{\tilde t}_R}}$. The loop functions
$P_{M_{\gamma, g}}^{LL(LR)}$ can be found in
Ref.\cite{Gabrielli:2004yi}. Finally, $U$ and $V$ are the matrices
that diagonalize chargino mass matrix.

Notice that the terms in $R_{M_{\gamma}}^{LL}$ and $R_{M_{g}}^{LL}$
which are enhanced by $m_{\chi_i}/m_b$ in Eq.(\ref{Rterms}) lead to
the large effects of chargino contributions to $C_{7\gamma}$ and
$C_{8g}$, respectively. Also the dependence of these terms on Yukawa
bottom $Y_b$ enhance the $LL$ contributions in $C_{7\gamma,8g}$ at
large $\tan\beta$. In the case of light stop-right, the function
$R_C^{RL}$ of the $Z$-penguin contribution is largely enhanced. In
order to understand the impact of the chargino contributions in
$B\to K \pi$ process, it is very useful to present the explicit
dependence of the Wilson coefficients $C_{7,9,7\gamma,8g}$ in terms
of the relevant mass insertions. For gaugino mass $M_2=200$ GeV,
squark masses $\tilde{m}=500$ GeV, light stop
$\tilde{m}_{\tilde{t}_R}=150$ GeV, $\mu = 400$ GeV, and $\tan \beta
=10$, we obtain \bea C_7^{\chi} &\simeq& 0.000002
(\delta^u_{LL})_{32} - 0.000011
(\delta^u_{RL})_{31} - 0.000046 (\delta^u_{RL})_{32},\\
C_9^{\chi} &\simeq&  0.00000039 (\delta^u_{LL})_{32} + 0.000037
(\delta^u_{RL})_{31} + 0.000165 (\delta^u_{RL})_{32},\\
C_{7\gamma}^{\chi} &\simeq& - 0.011 (\delta^u_{LL})_{31} - 0.05
(\delta^u_{LL})_{32} - 0.00043
(\delta^u_{RL})_{31} - 0.002 (\delta^u_{RL})_{32},\label{C7gamma}\\
 C_{8g}^{\chi} &\simeq& - 0.0032 (\delta^u_{LL})_{31} - 0.0014
(\delta^u_{LL})_{32} - 0.0003 (\delta^u_{RL})_{31} - 0.0012
(\delta^u_{RL})_{32}.
 \eea
From these results, it is clear that the Wilson coefficient
$C_{7\gamma}^{\chi}$ seems to give the dominant contribution,
specially through the $LL$ mass insertion. However, one should be
careful with this contribution since it is also the main
contribution to the $b\to s \gamma$, and stringent constraints on
$(\delta^u_{LL})_{32}$ are usually obtained, specially with large
$\tan \beta$. Finally, as expected from Eq.(\ref{Rterms}), only $LL$
contributions to $C^{\chi}_{7\gamma}$ and $C^{\chi}_{8g}$ have
strong depend on the value of $\tan \beta$. For instance with $\tan
\beta=40$, these contributions are enhanced with a factor 4, while
the result of $C^{\chi}_{7,9}$ and $LR$ part of $C^{\chi}_{7\gamma}$
and $C^{\chi}_{8g}$ change from the previous ones by less than
$2\%$.

\section{On the constraints from $BR(B\to X_s \gamma)$}
In this section we revise the constraints on SUSY flavor structure
which arise from the experimental measurements of the branching
ratio of the $B\to X_s \gamma$ \cite{bsgmeas}: \be 2 \times
10^{-4} < BR(b\to s \gamma) < 4.5 \times 10^{-4}~~~~ (\rm{at}~ 95
\% \rm{C.L.}).\label{bsg-exp} \ee In supersymmetric models, there
are additional contributions to $b\to s\gamma$ decay besides the
SM diagrams with $W$-gauge boson and an up quark in the loop. The
SUSY particles running in the loop are: charged Higgs bosons
($H^{\pm}$) or chargino with up quarks and gluino or neutralino
with down squarks. The total amplitude for this decay is sum of
all these contributions. As advocated in the introduction, the
neutralino contributions are quite small and can be safely
neglected. Also the charged Higgs contributions are only relevant
at very large $\tan \beta$ and small charged Higgs mass.
Therefore, we consider chargino and gluino contributions only to
analyze the possible constraints on the mass insertions
$(\delta^u_{AB})_{32}$ and $(\delta^d_{AB})_{23}$, where $A\equiv
L,R$.

Although the gluino contribution to $b\to s \gamma$ is typically
very small in models with minimal flavor structure, it is
significantly enhanced in models with non minimal favor structure
\cite{Gabrielli:2000hz}. In this class of models, both chargino
and gluino exchanges give large contribution to the amplitude of
$b\to s \gamma$ decay, and hence, they have to be simultaneously
considered in analyzing the constraints of the branching ratio
$BR(b\to s \gamma)$.

The relevant operators for this process are $Q_2$, $Q_{7\gamma}$,
and $Q_{8g}$. The contributions of the other operators in
Eq.(\ref{Heff}) can be neglected. The branching ratio $BR(b\to s
\gamma)$, conventionally normalized to the semileptonic branching
ratio $BR^{exp}(B\rightarrow X_c e \nu)=(10.4\pm0.4)\%$
~\cite{PDG}, is given by~\cite{bsgNLO} \bea BR^{\rm NLO}(B\to
X_s\gamma) &=&BR^{\rm{exp}}(B\rightarrow X_c e \nu )
\frac{|V_{ts}^{*} V_{tb}|^2}{|V_{cb}|^2} \frac{6 \alpha_{em}}{\pi
g(z) k(z)}\left(1-\frac{8}{3}\frac{\alpha_s (m_b)}{\pi}\right) \no
&\times& \left(|D|^2+A\right)(1+\delta_{np})~, \label{BRSM} \eea
with \bea D&=& C_7^{(0)}(\mu)+\frac{\alpha_s (\mu)}{4\pi}
\left(C_7^{(1)}(\mu)+ \sum_{i=1}^{8}C_i^{(0)}(\mu)
\left[r_i(z)+\gamma_{i7}^{(0)}
\log{\frac{m_b}{\mu}}\right]\right), \no A&=& \left(
e^{-\alpha_s(\mu) \log{\delta (7+2\log{\delta})/3\pi}}-1\right)
|C_7^{(0)}(\mu )|^2 +\frac{\alpha_s (\mu)}{\pi}\sum_{i\le
j=1}^{8}C_i^{(0)}(\mu ) C_j^{(0)}(\mu ) f_{ij}(\delta)~,\nonumber
\eea where $z=m_c^2/m_b^2$, $\mu$ is the renormalization scale
which is chosen of order $m_b$, and $\rho$ is photon energy
resolution. The expressions for $C_i^{(0)}$, $C_i^{(1)}$, and the
anomalous dimension matrix $\gamma$, together with the functions
$g(z)$, $k(z)$, $r_i(z)$ and $f_{ij}(\delta)$, can be found in
Ref.~\cite{bsgNLO}.  The term $\delta_{np}$ (of order a few
percent) includes the non-perturbative $1/m_b$ \cite{bsgNPmb} and
$1/m_c$ \cite{bsgNPmc} corrections. From the formula above we
obtain the theoretical result for $BR(B\to X_s \gamma)$ in the SM
which is given by \be BR^{\rm NLO}(B\to X_s\gamma)=(3.29\pm
0.33)\times 10^{-4} \label{bsgSM} \ee where the main theoretical
uncertainty comes from uncertainties in the SM input parameters,
namely $m_t,~\alpha_s(M_Z), ~\alpha_{em},~m_c/m_b,~m_b, V_{ij}$,
and the small residual scale dependence. The central value in
Eq.(\ref{bsgSM}) corresponds to the following central values for
the SM parameters $m_t^{\rm pole} \simeq m_t^{\rm
\overline{MS}}(m_Z) \simeq 174\,\rm GeV$, $m_b^{\rm pole} =
4.8\,\rm GeV$, $m_c^{\rm pole} = 1.3\,\rm GeV$, $\mu = m_b$,
$\alpha_s(m_Z) = 0.118$, $\alpha_e^{-1}(m_Z) = 128$,
$\sin^2\theta_W = 0.23$ and a photon energy resolution
corresponding to $\rho=0.9$ is assumed.

The SUSY contributions to the Wilson coefficients $C_{7\gamma}$ and
$C_8g$ at leading order are given in the previous section. In
general, the SUSY effects in $b\to s \gamma$ decay can be
parameterized by introducing $R_{7,8}$ and $\tilde{R}_{7,8}$
parameters defined at the electroweak scale as \be
R_{7,8}=\frac{\left(C_{7\gamma,8g}-C^{SM}_{7\gamma,8g}\right)}
{C_{7\gamma,8g}^{SM}},~~~
\tilde{R}_{7,8}=\frac{\tilde{C}_{7\gamma,8g}}{C_{7\gamma,8g}^{SM}},
\label{R78} \ee where $C_{7\gamma,8g}$ include the total
contribution while $C_{7\gamma,8g}^{SM}$ contains only the SM ones.
Note that in $\tilde{C}_{7\gamma,8g}$, which are the corresponding
Wilson coefficients for $\tilde{Q}_{7\gamma,8g}$ respectively, we
have set to zero the SM contribution. Inserting these definitions
into the $BR(B\to X_s\gamma$) formula in Eq.(\ref{BRSM}) yields a
general parametrization of the branching ratio
\cite{Gabrielli:2000hz,kagan}\bea {\rm BR}(B\to X_s\gamma) &=&
{\rm BR}^{\rm{SM}}(B\to X_s\gamma)\left(1 + 0.681
Re\left(R_7\right) + 0.116\left[\vert R_7\vert ^2+ \vert
\tilde{R}_7\vert^2\right]
\right.\nonumber \\
&+& \left.0.083 Re(R_8) + 0.025 \left[Re(R_7 R_8^*) +
Re(\tilde{R}_7 \tilde{R}_8^*)\right] \right.\nonumber \\
&+& \left.0.0045\left[\vert
R_8\vert^2+\vert\tilde{R}_8\vert^2\right]\right)~. \label{bsgPAR}
\eea From this parametrization, it is clear that $C_{7\gamma}$
would give the dominant new contribution (beyond the SM one) to
the $BR(B\to X_s \gamma)$. Using the allowed experimental range
given in Eq.(\ref{bsg-exp}), one can impose stringent constraints
on $C_{7\gamma}$, and hence on the corresponding mass insertions.
It is also remarkable that $R_7$ and $\tilde{R}_7$ have different
contributions to the $BR(B\to X_s \gamma)$, therefore, the
possible constraints on $C_{7\gamma}$ and hence on the LL and LR
mass insertions would be different from the constraints on
$\tilde{C}_{7\gamma}$ and hence on the RR and RL mass insertions,
unlike what has been assumed in the literatures. Furthermore,
since the leading contribution to the branching ratio is due to
$Re(R_7)$, the CP violating phase of $C_{7\gamma}$ will play a
crucial role in the possible constraints imposed by $BR(B\to X_s
\gamma)$.

Note that the constraints obtained in Ref.~\cite{ggms}, namely
$(\delta_{LR}^d)_{23} \leq 1.6\times 10^{-2}$ and
$(\delta_{LL}^d)_{23}$ is unconstrained are based on the assumption
that the gluino amplitude is the dominant contribution to $b\to
s\gamma$, even dominant with respect to the SM amplitude. Although
this a very acceptable assumption in order to derive a conservative
constraints on the relevant mass insertions, it is unrealistic and
usually lead to unuseful constraint. The aim of this section is to
provide a complete analysis of the $b\to s\gamma$ constraints by
including the SM, chargino and gluino contributions.

Let us start first with gluino contribution as the dominant SUSY
effect to $b\to s \gamma$ decay. We assume that the average squark
mass of order $500$ GeV and we consider three representative
values for $x = (m_{\tilde{g}}/m_{\tilde{q}})^2=0.3, 1$, and $4$.
We also assume that the SM value for $BR(B\to X_s \gamma)$ is
given by $3.29 \times 10^{-4}$, which is the central value of the
results in Eq.(\ref{bsgSM}). In these cases we find that both the
mass insertions $\vert (\delta^d_{LL})_{23}\vert$ and $\vert
(\delta^d_{RR})_{23}\vert$ are unconstrained by the branching
ratio of $b\to s \gamma$ for any values of their phases. The upper
bounds on $\vert (\delta^d_{LR})_{23}\vert$ and $\vert
(\delta^d_{RL})_{23}\vert$ from $b\to s \gamma$ decay are give in
Table \ref{table3}.
\begin{table}[t]
\begin{center}
\begin{tabular}{|c|c|c|}
\hline x & $\vert (\delta^d_{LR})_{23}\vert$ & $\vert
(\delta^d_{RL})_{23}\vert$\\
\hline
    & (a) 0.0116 &   \\
0.3& (b) 0.0038 & 0.0038 \\
    & (c) 0.0012 & \\
\hline
    & (a) 0.02 &   \\
 1  & (b) 0.006 & 0.006 \\
    & (c) 0.002 & \\
\hline
    & (a) 0.006 &   \\
 4  & (b) 0.015 & 0.016 \\
    & (c) 0.0045 & \\
\hline
\end{tabular}
\end{center}
\caption{Upper bounds of $\vert (\delta^d_{LR(RL)})_{23}\vert$
from $b\to s \gamma$ decay for $m_{\tilde{q}}=500$ GeV and
$\rm{arg}(\delta^d_{LR(RL)})_{23}=0~(a), \pi/2~(b), \pi~(c)$
respectively.} \label{table3}
\end{table}
As can be seen from these results, the limits on $\vert
(\delta^d_{LR})_{23}\vert$ are quite sensitive to the phase of
this mass insertion, unlike the bounds on $\vert
(\delta^d_{RL})_{23}\vert$. Also, as suggested by
Eq.(\ref{bsgPAR}), the bounds on LR coincides with the ones on RL
only if $\rm{arg}(\delta^d_{LR(RL)})_{23}= \pi/2$. Note that in
this case $Re(R_7)$ vanishes and the expression of the branching
ratio is a symmetric under exchange $R_7$ and $\tilde{R}_7$.

Now we consider the chargino contribution as the dominant SUSY
effect to $b\to s\gamma$ in order to analyze the bounds on the
relevant mass insertions in the up squark sector. From the
expression of $C^{\chi}_{7 \gamma}$ in Eq.(\ref{C7gamma}), which
provide the leading contribution to the branching ratio of $b\to
s\gamma$, it is clear that one can derive strong constraints on
$(\delta^u_{LL})_{32}$ and $(\delta^u_{LL})_{31}$ and a much
weaker constraints (essentially no constrain) on
$(\delta^u_{RL})_{32}$ and $(\delta^u_{RL})_{31}$. The resulting
bounds on $(\delta^u_{LL})_{32}$ and $(\delta^u_{LL})_{31}$ as
functions of the gaugino mass $M_2$ and the average squark mass
$\tilde{m}$ are presented in Tables 5, for $\tan \beta =10$ and
$\mu =400$ GeV.

\begin{table}[t]
\begin{tabular}{|c|c|c|c|c|}
\hline $M_2 \backslash m$ & 300 & 500 & 700 & 900\\
\hline
   & (a) 0.04&  0.065 &  0.095 & 0.14  \\
150& (b) 0.14 & 0.24 & 0.37 & 0.54 \\
 & (c) 0.51 & 0.85 & ------& ------\\
\hline
    & (a) 0.053&  0.075 & 0.1 & 0.15  \\
250& (b) 0.20 & 0.28 & 0.4 & 0.55\\
    & (c) 0.70 & ------ & ------ & ------\\
\hline
    & (a) 0.07&  0.09 &  0.12 & 0.16  \\
350& (b) 0.26 & 0.33 & 0.45 & 0.6 \\
    & (c) 0.92 & ------ & ------ & ------\\
\hline
    & (a) 0.085&  0.105 &  0.14 & 0.16  \\
450& (b) 0.33 & 0.4 & 0.5 & 0.6\\
    & (c) ------ & ------ & ------&------ \\
\hline
\end{tabular}
~~~~ \begin{tabular}{|c|c|c|c|c|}
\hline $M_2 \backslash m$ & 300 & 500 & 700 & 900\\
\hline
    & (a) 0.17&  0.28 &  0.45 & 0.65  \\
150& (b) 0.65 & ------ & ------ & ------ \\
    & (c) ------ & ------ & ------& ------\\
\hline
    & (a) 0.24 &  0.34 & 0.48 & 0.67  \\
250&(b) 0.86 & ------ & ------ & ------\\
    & (c) ------ & ------ & ------ & ------\\
\hline
    & (a) 0.32&  0.4 &  0.52 & 0.73  \\
350&(b) ------ & ------ & ------ & ------ \\
    & (c) ------ & ------ & ------ & ------\\
\hline
    & (a) 0.45 &  0.48 &  0.62 & 0.8  \\
450&(b) ------ & ------ & ------ & ------\\
    & (c) ------ & ------ & ------&------ \\
\hline
\end{tabular}
\caption{Upper bounds of $\vert (\delta^u_{LL})_{32}\vert$ (left)
and $\vert (\delta^u_{LL})_{31}\vert$ (right) from $b\to s \gamma$
decay for $\tan \beta=10$ and $\mu=400$ GeV and
$\rm{arg}(\delta^u_{LL})_{32(31)}= 0~ (a), \pi/2~ (b), \pi~ (c)$
respectively.} \label{table5}
\end{table}

The results in Table 5 correspond to positive sign of $\mu$. If one
assumed negative sign of $\mu$, the constraints on $\vert
(\delta^u_{LL})_{32}\vert$ and $\vert (\delta^u_{LL})_{31}\vert$
with $\rm{arg}(\delta^u_{LR})_{32(31)}$ will be exchanged with the
corresponding ones with $\rm{arg}(\delta^u_{LL})_{32(31)} + \pi$.
Thus, in Table 5, the results of case $(a)$ will be replaced with
the results of $(c)$ and vice versa. For larger values of $\tan
\beta$, the above constraints will be reduced by the factor
$\left(\tan\beta/10\right)$. Note also that, because of the $SU(2)$
gauge invariance the soft scalar mass $M_{Q}^2$ is common for the up
and down sectors. Therefore, one gets the following relations
between the up and down mass insertions \be (\delta^d_{LL})_{ij} =
\left[V_{CKM}^+~ (\delta^u_{LL})~ V_{CKM} \right]_{ij}~. \ee Hence,
\be (\delta^d_{LL})_{32} =(\delta^u_{LL})_{32} +
\mathcal{O}(\lambda^2)~. \ee As a result, the constraints obtained
from the chargino contribution to $b \to s \gamma$ transition on
$\vert(\delta^u_{LL})_{32}\vert$ can be conveyed to a constraint on
$\vert(\delta^d_{LL})_{32}\vert$ which equals to
$\vert(\delta^d_{LL})_{23}\vert$, due to the hermiticity of
$(M_{D}^2)_{LL}$. This is the strongest constraint one may obtain on
$\vert(\delta^d_{LL})_{23}\vert$, and therefore it should be taken
into account in analyzing the LL part of the gluino contribution to
the $b \to s$.

Finally we consider the scenario in which both gluino and chargino
exchanges are assumed to contribute to $b \to s \gamma$
simultaneously with relevant mass insertions, namely
$(\delta^d_{LR})_{23}$ and $(\delta^u_{LL})_{32}$. It is known
that these two contributions could give rise to a substantial
destructive or constructive interference with the SM amplitude,
depending on the relative sign of these amplitudes. Recall that in
minimal supersymmetric standard model with the universality
assumptions, the gluino amplitude is negligible, since
$(\delta^d_{LR})_{23} \lsim \mathcal{O}(10^{-6})$, and the
chargino contribution at large $\tan \beta$ is the only relevant
SUSY contribution. In this class of model, depending on the sign
of $\mu$ the chargino contribution gives destructive interference
with the SM result.

In generic SUSY model, the situation is different and the
experimental results of the branching ratio of $b \to s \gamma$
can be easily accommodated by any one of these contributions. Also
since the gluino and the chargino contributions are given in terms
of the parameters of the up and down squark sectors, they are, in
principle, independent and could have destructive interference
between themselves or with the SM contribution. We stress that we
are not interested in any fine tuning region of the parameter
space that may lead to a large cancelation. We are rather
considering the general scenario with large down and up mass
insertions favored by the CP asymmetries of different $B$
processes. In this case, both gluino and chargino contributions to
$b\to s\gamma$ are large and cancelation of order $20-50 \%$ can
take place.

\begin{figure}
\begin{center}
\includegraphics[width=10.5cm]{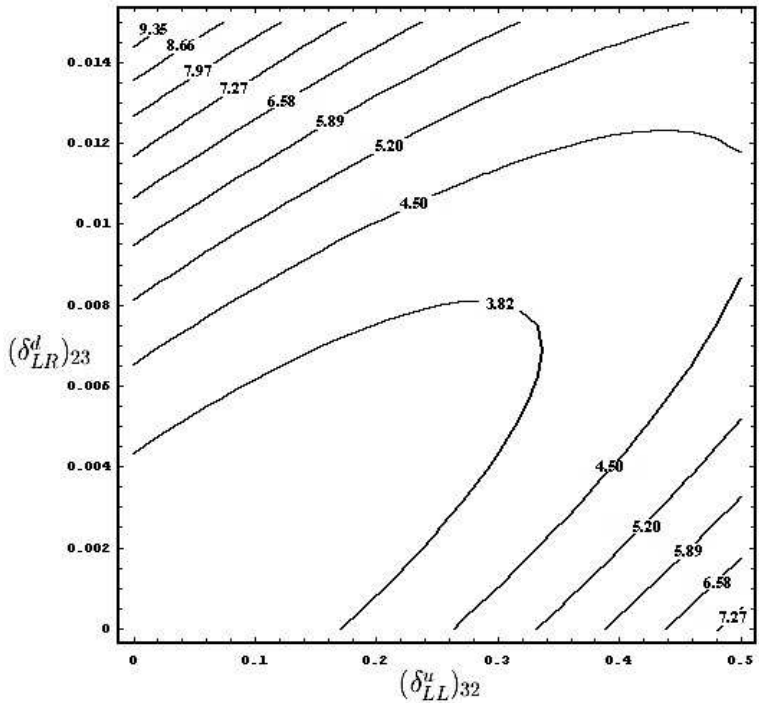}
\caption{Contour plot for $BR(b\to s \gamma)\times 10^{4}$ as
function of $(\delta^{d}_{LR})_{23}$ and $(\delta^u_{LL})_{32}$. }
\vspace{-0.6cm}
\end{center}\end{figure}

Now, it is clear that the previous constrained obtained on
$(\delta^d_{LR})_{23}$ and $(\delta^u_{LL})_{32}$ in Tables 4 and
5 will be relaxed. We plot the corresponding results for the
correlations between $(\delta^d_{LR})_{23}$ and
$(\delta^u_{LL})_{32}$ in Fig. 1. Here we consider the relation
$(\delta^d_{LL})_{23} = (\delta^u_{LL})_{32}$ into account and
also set $(\delta^u_{RL})_{32}$ to zero. The phases of
$(\delta^d_{LR})_{23}$ and $(\delta^u_{LL})_{32}$ are assumed to
be of order $\pi/2$ as favored by the CP asymmetry of $B\to \phi
K$. From this plot, we can see that constraints on these mass
insertions, particularly $(\delta^u_{LL})_{32}$ are relaxed.

\section{SUSY solution to the $R_c- R_n$ Puzzle}
Now we analyze the supersymmetric contributions to the $B\to K
\pi$ branching ratio. We will show that the simultaneous
contributions from penguin diagrams with chargino and gluino in
the loop could lead to a possible solution  to the $R_c-R_n$
puzzle. As mentioned in section 3, these penguin contributions
have three possible sources of large SUSY contribution to $B \to K
\pi$ processes:
\begin{enumerate}
\item Gluino mass enhanced $O_{7\gamma}$ and $O_{8g}$ which depend on $(\delta^d_{LR})_{23}$
and $(\delta^d_{RL})_{23}$.

\item Chargino mass enhanced $O_{7\gamma}$ and $O_{8g}$ which depend on $\tan \beta
(\delta^u_{LL})_{23}$.

\item Right handed stop mass enhanced $Z$ penguin which is given in terms of
$(\delta^u_{RL})_{32}$. \end{enumerate}

For the same inputs of SUSY parameters that we used above:
$m_{\tilde{g}}=500$ GeV, $m_{\tilde{q}}=500$ GeV,
$m_{\tilde{t}_R}=150$ GeV, $M_2=200$ GeV, $\mu = 400$ GeV, and
$\tan \beta=10$, one finds the following SUSY contributions to the
amplitudes of $B\to K \pi$ \bea A_{\bar{B}^0\to \pi^0 \bar{K}^0}
\times 10^{7} &\simeq & -9.82~i~\left[(\delta^d_{LR})_{23}
+(\delta^d_{RL})_{23}\right]+ 0.036~i~(\delta^u_{LL})_{32} -
0.02~i~(\delta^u_{RL})_{32},\no A_{\bar{B}^0\to \pi^+ K^-} \times
10^{7} &\simeq & 14.04~i~\left[(\delta^d_{LR})_{23}
+(\delta^d_{RL})_{23}\right]+ 0.06~i~(\delta^u_{LL})_{32}-
0.001~i~(\delta^u_{RL})_{32},\no A_{B^-\to \pi^0 K^-} \times
10^{7} &\simeq & 9.9~i~\left[(\delta^d_{LR})_{23}
+(\delta^d_{RL})_{23}\right] - 0.04~i~(\delta^u_{LL})_{32} +
0.024~i~(\delta^u_{RL})_{32},\no A_{B^-\to \pi^- K^0} \times
10^{7} &\simeq & 13.89~i~\left[(\delta^d_{LR})_{23}
+(\delta^d_{RL})_{23}\right] +
0.05~i~(\delta^u_{LL})_{32}-0.006~i~(\delta^u_{RL})_{32}.\nonumber
\eea

It is remarkable that for the amplitudes $A_{\bar{B}^0\to \pi^0
\bar{K}^0}$ and $A_{B^-\to \pi^0 K^-}$, which suffer from a large
discrepancy between their SM values and their experimental
measurements, the SUSY contributions have the following features:
$(i)$ the effect of $(\delta^u_{RL})_{32}$ is not negligible as in
the other amplitudes, $(ii)$ there can be a distractive
interference between the $(\delta^d_{LR})_{23}$ and
$(\delta^u_{LL})_{32}$ contributions. As we will see below, these
two points are important in saturating the experimental results by
supersymmetry. Also note that the effect of gluino contribution
through $O_{7\gamma}$ is very small and the contribution of
$(\delta^d_{LR})_{23}$ is mainly due to $O_{8g}$. However the
chargino effect of $O_{7\gamma}$ can be enhanced by $\tan\beta$.

\begin{figure}
\begin{center}
\includegraphics[width=9.5cm]{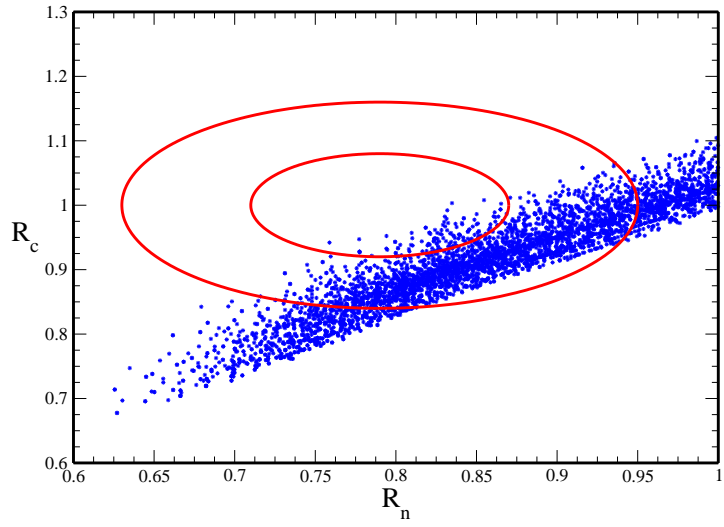}
\caption{$R_c-R_n$ correlation in SUSY models with
$|\duLRtwo|\simeq 1$, $|\ddLR|\in [0.001,0.01]$ and $|\duLLtwo|\in
[0.1,1]$; see the text for the other parameters. The small and
large ellipses correspond to $1\sigma$ and $2\sigma$ experimental
results, respectively.}\label{RnRc.eps} \vspace{-0.6cm}
\end{center}\end{figure}

We present our numerical results for the correlation between the
total contributions (SM+SUSY) to the $R_n$ and $R_c$ in Fig.
\ref{RnRc.eps}. We have scanned over the relevant mass insertions:
$(\delta^u_{LL})_{32}$, $(\delta^d_{LR})_{23}$ and
$(\delta^u_{RL})_{32}$, since we have assumed
$(\delta^u_{LL})_{32} \simeq (\delta^d_{LL})_{23}$ and
$(\delta^d_{LR})_{23}\simeq (\delta^d_{RL})_{23}$. We considered
$\vert (\delta^u_{LL})_{32} \vert \in [0.1,1]$, $\vert
(\delta^d_{LR})_{23} \vert \in [0.001,0.01]$,
$\arg[(\delta^u_{LL})_{32}] \in
[-\pi,\pi]$,$\arg[(\delta^d_{LR})_{23}] \simeq \pi/3$ (which is
preferred by $S_{\phi K_S}$), and $(\delta^u_{RL})_{32}]=1$ (in
order to maximize the difference between $R_n$ and $R_c$). As can
be seen from the results in Fig. \ref{RnRc.eps}, the experimental
results of $R_n$ and $R_c$ at $2 \sigma$ can be naturally
accommodated by the SUSY contributions. However, the results at $1
\sigma$ can be only obtained by a smaller region of parameter
space. In fact, the values of $R_c$ is predicted to be less than
one for the most of the parameter space. Therefore, it will be
nice accordance with SUSY results if the experimental result of
$R_c$ goes down.

In order to understand the results in Fig. \ref{RnRc.eps} and the
impact of the SUSY on the correlation between $R_n$ and $R_c$, we
extend the parametrization introduced in section 2 for the relevant
amplitudes by including the SUSY contribution \cite{Khalil:2004yb}.
In this case, Eqs.(\ref{paramerization1}) can be written as \bea
A_{B^-\to\pmkz}&=&\lambda_c A_{\pi \bar{K}} P\left[e^{i\theta_P}+r_A
e^{i\delta_A}e^{-i\gamma}\right]
\label{susypar1} \\
\sqrt{2}A_{B^-\to\pzkm}&=&\lambda_c A_{\pi \bar{K}}
P\big[e^{i\theta_P}+ \left(r_A e^{i\delta_A} + r_C
e^{i\delta_C}\right) e^{-i\gamma}+ r_{EW} e^{i\theta_{EW}}
e^{i\delta_{EW}}
\big] \label{susypar2}\\
A_{\overline{B}^0\to\ppkm}&=&
\lambda_cA_{\pi\Kbar}P\left[e^{i\theta_P}+\left(r_A e^{i\delta_A}+
r_T
e^{i\delta_T}\right)e^{-i\gamma} + r_{EW}^C e^{i\theta_{EW}}e^{i\delta_{EW}^C}\right], \label{susypar3}\\
-\sqrt{2}A_{\overline{B}^0\to\pzkz}&\!=\!&\lambda_cA_{\pi\Kbar}P\Big[e^{i\theta_P}\!+\!
\left( r_A e^{i\delta_A} + r_T e^{i\delta_T} - r_C
e^{i\theta_C}e^{i\delta_C}\right) e^{-i\gamma} + r_{EW}^C
e^{i\theta_{EW}^C}e^{i\delta_{EW}^C}\nonumber\\ &-& r_{EW}
e^{i\theta_{EW}}e^{i\delta_{EW}}\Big]. \label{susypar4} \eea The
parameters $\delta_A, \delta_C, \delta_T, \delta_{EW},
\delta_{EW}^C$ and $\theta_P, \theta_{EW}, \theta_{EW}^C$ are the
CP conserving (strong) and the CP violating phase, respectively.
Note that the parameters $P, r_{EW}, r_{EW}^C$ are now defined as
\bea P e^{i \theta_P}
e^{i\delta_P}&=&\alpha_4^c-\frac{1}{2}\alpha^c_{4,EW}+\beta_3^c+\beta^c_{3,EW}
,\no r_{EW} e^{i \theta_{EW}}
e^{i\delta_{EW}}&=&\left[\frac{3}{2}(R_{K\pi}\alpha^c_{3,EW}+
\alpha^c_{4,EW})\right]/P ,\no
 r^C_{EW} e^{i
\theta_{EW}^C}e^{i\delta_{EW}^C}&=&
\left[\frac{3}{2}(\alpha^c_{4,EW}-\beta^c_{3,EW})\right]/P .\eea

First let us include some assumptions to simplify our formulae. As
is mentioned before, $\alpha_{4}^p, \alpha_{3,EW}^p,
\alpha_{4,EW}^p, \beta_{3}^p, \beta_{3,EW}^p, \beta_{4,EW}^p$
receive SUSY contributions through the Wilson coefficients. The
upper index $p$ takes both $u$ and $c$, however the contribution
with $u$ index is always suppressed by the factor
$\epsilon_{KM}\simeq 0.018$ so that its SUSY contributions can be
safely neglected comparing to the one with the index $c$. As a
result, $(r_A e^{i\delta_A}), (r_{C} e^{i\delta_C})$ and $(r_T
e^{i\delta_T})$ receive a correction of a factor
$1/|1+\frac{P^{\susy}}{P^{\sm}}|$.

Secondly we assume that the strong phase for SM and SUSY are the
same. We found that this is a reasonable assumption in QCD
approximation in which the main source of the strong phase comes
from hard spectator and weak annihilation diagrams. This leads us
to the following parametrization: \bea
P e^{i\delta_P}e^{i\theta_P}&=& P^{\sm}e^{i\delta_P}(1+ke^{i\theta^{\prime}_P}) \\
r_{EW} e^{i\delta_{EW}}e^{i\theta_{EW}}&=&
(r_{EW})^{\sm}e^{i\delta_{EW}}
(1+le^{i\theta^{\prime}_{EW}}) \label{rEW}\\
r_{EW}^C e^{i\delta_{EW}^C}e^{i\theta_{EW}^C}&=&
(r_{EW}^{C})^{\sm}e^{i\delta_{EW}^C}(1+me^{i\theta^{C^\prime_{EW}}}).\label{rEWC}
\eea where \bea k e^{i\theta^{\prime}_P}&\equiv&
\frac{(\alpha_4^c-\frac{1}{2}\alpha^c_{4,EW}+\beta_3^c+\beta^c_{3,EW}
)_{\susy}}{(\alpha_4^c-\frac{1}{2}\alpha^c_{4,EW}+\beta_3^c+\beta^c_{3,EW}
)_{\sm}}, \\
le^{i\theta^{\prime}_q}&\equiv&
\frac{(R_{K\pi}\alpha^c_{3,EW}+\alpha^c_{4,EW})_{\susy}}{(R_{K\pi}\alpha^c_{3,EW}+
\alpha^c_{4,EW})_{\sm}},\\
me^{i\theta^{\prime}_{q_C}} &\equiv&
\frac{(\alpha^c_{4,EW}-\beta^c_{3,EW})
_{\susy}}{(\alpha^c_{4,EW}-\beta^c_{3,EW})_{\sm}} \eea The index
SM (SUSY) mean to keep only SM (SUSY) Wilson coefficients in
 $\alpha_{i (,EW)}^p$ and $\beta_{i (,EW)}^p$. Using these parameters, we also have
 \be
r_A e^{i\delta_A}=\frac{(r_A e^{i\delta_A})_{\sm}}{\left|1+k
e^{i\theta^{\prime}_P}\right|}, \ \ r_C e^{i\delta_C}=\frac{(r_C
e^{i\delta_C})_{\sm}}{\left|1+k e^{i\theta^{\prime}_P}\right|}, \
\ r_T e^{i\delta_T}=\frac{(r_T e^{i\delta_T})_{\sm}}{\left|1+k
e^{i\theta^{\prime}_P}\right|} \label{risusy}\ee

Now let us investigate the $R_c-R_n$ puzzle. We shall follow the standard procedure; to simplify and
expand the formulae. Considering the numbers obtained above, we shall simplify our formulae
by assuming
\begin{enumerate}
\item the strong phases are negligible, i.e., $\delta_P, \delta_A, \delta_C,\delta_{EW},
\delta_{EW}^C$ are all zero.
\item the annihilation tree contribution is  negligible, i.e. $r_A \simeq 0$
\item the color suppressed tree contribution is negligible, i.e. $r_C e^{i\delta_C}
\sim r_T e^{i\delta_T}$.
\end{enumerate}
Using these assumptions,  we expand $R_c, R_n$ and $R_c-R_n$. We
expand in terms of $r_T$ and $r_{EW}$ and $r_{EW}^C$ up to the
second order. As a result, we obtain \bea
&&\hspace{-1.1cm}R_c\simeq 1\!+\!r_T^2\!-\!2r_T\cos
(\gamma\!+\!\theta_P)\!+\!2r_{EW}\cos
(\theta_P\!-\!\theta_{EW})\!-\!2r_T r_{EW}\cos (\gamma
\!+\!\theta_{EW})
\\
&&\hspace{-1.1cm}R_c-R_n \simeq 2r_T r_{EW}\cos (\gamma
+2\theta_P-\theta_{EW}) -2r_T r_{EW}^C\cos (\gamma
+2\theta_{EW}-\theta_{EW}^{C})
\label{eq:RcRn} \eea Now, let us find the configuration which lead
to $R_c-R_n>0.2$. Looking at Eq. (\ref{eq:RcRn}), we can find that
in general, the larger the values of $r_T$, $r_{EW}$ and $r_{EW}^C$
are, the larger the splitting between $R_c$ and $R_n$ we would
acquire. The phase combinations $\theta_P-\theta_{EW}$ and
$\theta_P+\gamma$ also play an important role. The possible solution
of $R_c-R_n$ puzzle by enhancing $r_{EW}$, which we parameterize as
$l$, has been intensively studied in the literature
\cite{kpi-puzzle}. As we will see in the following, $r_T$ can also
be enhanced due to the factor $ke^{i\theta_P^{\prime}}$ which
contributes destructively against the SM and diminish $P$. However,
since $P$ is the dominant contribution to the $B\to K\pi$ process,
the branching ratio is very sensitive to $ke^{i\theta_P^{\prime}}$.
Therefore, we are allowed to vary $ke^{i\theta_P^{\prime}}$ only in
a range of the theoretical uncertainty of QCD factorization, which
gives about right sizes of the $B\to K\pi$ branching ratios. As
showed in Ref.\cite{Khalil:2004yb}, we would be able to reduce $P$
at most by 30 \%, which can be easily compensated by the error in
the transition form factor $F^{B\to\pi, K}$.

Considering the tiny effect from the second term in Eq.
(\ref{eq:RcRn}), in order to achieve $R_c-R_n\gsim 0.2$, we need
$r_{T} r_{EW}$ larger than about 0.1 or equivalently, $r_{EW}$
larger than about 0.5 with $r_T^{\sm}$. In
Ref.\cite{Khalil:2004yb}, it was emphasized that with $k=0$, one
needs $l\gsim 2$ to reproduce the experimental values while an
inclusion of a small amount of $k$ lowers this bound
significantly. For the SUSY parameters that we have considered
above, the following results for our SUSY parameters $k, l,$ and
$m$ are obtained \bea
ke^{i\theta_P}&=& -0.0019 \tan\beta \duLLtwo - 35.0 \ddLR +0.061\duLRtwo\label{eq:kSUSY}\\
le^{i\theta_q}&=& 0.0528 \tan\beta \duLLtwo-2.78\ddLR +1.11\duLRtwo \label{eq:lSUSY}\\
me^{i\theta_{q_C}}&=& 0.134 \tan\beta\duLLtwo + 26.4\ddLR +1.62
\duLRtwo \label{eq:mSUSY} \eea Note that we do not consider
$(\delta^d_{23})_{RL}$ here but it is the same as $\ddLR$ with an
opposite sign (see also \cite{Khalil:2003bi}). Let us first discuss
the contributions from a single mass insertion $\duLLtwo, \ddLR$ or
$\duLRtwo$ to $\{k, l, m\}$; keeping only one mass insertion and
switching off the other two. In this case, one finds that the
maximum value of $\{k,l,m\}$ with $|\duLRtwo|=1$ is
$\{k,l,m\}=\{0.061, 1.11, 1.62\}$. Thus, in this case where $k$ is
almost negligible, we would need $l\simeq 2$ to explain the
experimental data. We have a chance to enlarge the coefficients for
$\duLRtwo$ by, for instance, increasing the averaged squark mass
$\tilde{m}_{\tilde{q}}$. However, even if we choose
$\tilde{m}_{\tilde{q}}=5$ TeV, we find that $l$ is increased only by
20 to 30 \%. The maximum contributions from $\ddLR$ and $\duLLtwo$
are found to be $\{k, l, m\}=\{0.18, 0.014, 0.13\}$ and $\{0.0019,
0.053, 0.13\}$, which are far too small to explain the experimental
data. The coefficients for $\ddLR$ depend on the overall factor
$1/\tilde{m}_{\tilde{q}}$ and on also the variable of the loop
function $x=m_{\tilde{g}}/\tilde{m}_{\tilde{q}}$ and we found that
$m_{\tilde{g}}=\tilde{m}_{\tilde{q}}=250$ GeV can lead to  100 \%
increase. However, the value of $l$ is still too small to deviate
$R_c-R_n$ significantly. As a whole, it is extremely difficult to
have  $R_c-R_n\gsim 0.2$ from a single mass insertion contribution.

Let us try to combine two main contributions, $\ddLR$ and $\duLRtwo$
terms. Using the previous input parameters and including the $b\to
s\gamma$ constraint $|\ddLR|$, the maximum value is found to be
$\{k, l, m\}=\{0.24, 1.12, 1.48\}$. In this case, it is easy to
check that the experimental data are not reproduced very well
\cite{Khalil:2004yb}. As discussed above, for a large value of the
averaged squark masses, $l$ increases while $k$ decreases. On the
contrary, $k$ also depends on the ratio of gluino and squark masses.
Hence we need to optimize these masses so as to increase $k$ and $l$
simultaneously. For instance, with
$m_{\tilde{g}}\hspace*{-0.1cm}=\hspace*{-0.1cm}250$ GeV and
$\tilde{m}_{\tilde{q}}\hspace*{-0.1cm}=\hspace*{-0.1cm}1$ TeV, we
obtain $\{k, l, m\}\hspace*{-0.1cm}=\hspace*{-0.1cm}\{0.30, 1.36,
1.90\}$ which leads to a result within the experimental bounds of
$R_c$ and $R_n$. Finally, we consider the case with the three
non-zero mass insertions. The main feature of this scenario is that
we expect a relaxation of the constraints on
$|\tan\beta\times\duLLtwo|$ and $|\ddLR|$ from the cancelation
between  $\ddLR$ and $\duLLtwo$ contributions to $b\to s\gamma$.
Under this circumstance, we observe much larger $R_c-R_n$ for
various combination of the phases in this scenario.

\section{SUSY contributions to the CP asymmetry of $B\to K \pi$}
We start this section by summarizing our convention for CP
asymmetry in $B \to K \pi$ processes. The time dependent CP
asymmetry for $B \to K \pi$ can be described by \be
A_{K\pi}(t)=A_{K\pi} \cos(\Delta M_{B_d} t) + S_{K \pi}
\sin(\Delta M_{B_d})t, \ee where $A_{K\pi}$ and $S_{K\pi}$
represent the direct and the mixing CP asymmetry respectively and
they are given by \be A_{K\pi} = \frac{\vert
\bar{\rho}(K\pi)\vert^2 -1}{\vert \bar{\rho}(K \pi)\vert^2 + 1},
~~~~~~~~ S_{K\pi} = \frac{2 Im (\bar{\rho}(K\pi))}{\vert
\bar{\rho}(K \pi)\vert^2 + 1}, \label{asym_pi}\ee where
$\bar{\rho}(K\pi)= e^{-i \phi_B} \frac{\bar{A}(K\pi)}{A(K\pi)}$.
The phase $\phi_B$ is the phase of $M_{12}$, the $B^0 - \bar{B}^0$
mixing amplitude. The $A(K\pi)$ and $\bar{A}(K\pi)$ are the decay
amplitudes for $B^0$ and $\bar{B}^0$ to $K\pi$, respectively.

The SM predicts that the direct and mixing asymmetry of $B\to K \pi$
decay are given by \be S_{K \pi} = \sin 2 \beta, ~~~~~~~~ C_{K \pi}
=0. \ee

The recent measurements of the CP asymmetries in $B \to K \pi$,
reported in Table 1, show significant discrepancies with the SM
predictions.  As mentioned above, SUSY can affect the results of the
CP asymmetries in $B$ decay, due to the new source of CP violating
phases in the corresponding amplitude. Therefore, deviation on $CP$
asymmetries from the SM expectations can be sizeable, depending on
the relative magnitude of the SM and SUSY amplitudes. In this
respect, SUSY models with non-minimal flavor structure and new CP
violating phases in the squark mass matrices, can generate large
deviations in the $B\to K \pi$ asymmetry. In this section we present
and discuss our results for SUSY contributions to the direct and
mixing CP asymmetries in $B \to K \pi$.
\subsection{SUSY contributions to the direct CP asymmetry in $B \to K
\pi$}
Using the general parametrization of the decay amplitudes of $B
\to K \pi$ given in Eqs.(\ref{susypar1}-\ref{susypar4}), one can
write the direct CP asymmetries $A^{CP}_{K\pi}$ as follows: \bea
A^{CP}_{K^-\pi^+} &\simeq & 2 r_T \sin \delta_T \sin(\theta_P +
\gamma) +2 r_{EW}^C \sin \delta_{EW}^C \sin(\theta_P -
\theta_{EW}^c) - r_T^2 \sin 2\delta_T \sin2(\theta_P+\gamma)\no
&+& 2 r_T r^C_{EW} \sin(\delta^C_{EW} - \delta_T)
\sin(\theta^C_{EW}+\gamma) - 4 r_T r^C_{EW} \sin \delta^C_{EW}
\sin(\theta_P -\theta^C_{EW}) \cos \delta_T\no &&
\cos(\theta_P+\gamma) - 4 r_T r^C_{EW} \sin \delta_T
\sin(\theta_P +\gamma) \cos \delta^c_{EW} \cos(\theta_P - \theta^C_{EW}),\label{CP1}\\
A^{CP}_{K^0\pi^-} &\simeq & 2 r_A \sin \delta_A
\sin(\theta_P+\gamma), \label{CP2}\\
A^{CP}_{K^0\pi^0} &\simeq & 2 r_{EW}^C \sin \delta_{EW}^C
\sin(\theta_P- \theta_{EW}^C) - 2 r_{EW}\sin \delta_{EW}
\sin(\theta_P- \theta_{EW}),\label{CP3}\\
A^{CP}_{K^-\pi^0} &\simeq & 2 r_{T} \sin \delta_{T}
\sin(\theta_P+\gamma) - 2 r_{EW}\sin \delta_{EW} \sin(\theta_P-
\theta_{EW})- r_T^2 \sin 2\delta_T \sin2(\theta_P+\gamma)\no &-& 2
r_T r_{EW} \sin(\delta_{EW} - \delta_T) \sin(\theta_{EW}+\gamma) -
4 r_T r_{EW} \sin \delta_{EW} \sin(\theta_P -\theta_{EW}) \cos
\delta_T\no && \cos(\theta_P+\gamma) - 4 r_T r_{EW} \sin \delta_T
\sin(\theta_P +\gamma) \cos \delta_{EW} \cos(\theta_P -
\theta_{EW}).\label{CP4} \eea From these expressions, it is clear
that if we ignore the strong phases, then the direct CP
asymmetries would vanish. However, Belle and BaBar collaborations
observed non-zero values for the $A^{CP}_{K\pi}$, thus we should
consider non-vanishing strong phases in this analysis. It is also
remarkable that the leading contributions to the direct CP
asymmetries are given by the linear terms of $r_i \equiv r_T, r_A,
r_{EW}, r_{EW}^C$, unlike the difference $R_c-R_n$ which receives
corrections of order $r_i r_j$. As in the previous section, we
have assumed that the color suppressed contributions are
negligible {\it i.e.}, $r_C e^{i\delta_C} = r_T e^{i\delta_T}$ and
we have neglected terms of order $r_i^2$ except for $r_T$ which is
typically larger than $r_{EW}$, $r_{EW}^C$, and $r_A$.

The rescattering effects parameterized by $r_A$ are quite small
$\left(r_A^{SM} \simeq \mathcal{O}(0.01)\right)$ therefore the CP
asymmetry in the decays $B^{\pm} \to K^0 \pi^{\pm}$ is expected to
be very small as can be easily seen from Eq.(\ref{CP2}). This
result is consistent with the experimental measurements reported
in Table 1. The sign of this asymmetry will depend on the relative
sign of $\sin \delta_A$ and $\sin(\theta_P+\gamma)$. Note that the
value of the angle $\gamma$ is fixed by the CP asymmetry in $B \to
\pi \pi$ to be of order $\pi/3$. The angle $\theta_P$ can also be
determined from the CP asymmetry $S_{\phi (\eta') K}$.

In the SM, the parameters $r_A, r^C_{EW}$ are much smaller than
$r_T, r_{EW}$ and $\theta_P=0$, therefore the following relation
among the direct CP asymmetries $A^{CP}_{K\pi}$ is obtained
$$ A^{CP}_{K^-\pi0} \gsim A^{CP}_{K^- \pi^+} \gsim
A^{CP}_{K^0\pi^0} > A^{CP}_{K^0 \pi^-}.$$ This relation is in
agreement with the numerical results listed in Table 3 for the
direct CP asymmetries in the SM with $\rho_{A,H}, \phi_{A,H}\simeq
1$. To change this relation among the CP asymmetries and to get
consistent correlations with experimental measurements, one should
enhance the electroweak penguin contributions to $\bar{B}^0 \to
K^- \pi^+$ decay amplitude, parameterized by $r^C_{EW}$.
Furthermore, a non-vanishing value of $\theta_P$, which is also
required to account for the recent measurements of $S_{\phi K_S}$
and $S_{\eta' K_S}$, is favored in order to obtain $A^{CP}_{K^-
\pi^+}>A^{CP}_{K^- \pi^0}$. It is worth mentioning that in the SM
and due to the fact that $\theta_P=0$ the second term in
Eq.(\ref{CP1}) and Eq.(\ref{CP4}) give destructive and
constructive interferences respectively with first terms. Thus one
finds $A^{CP}_{K^- \pi^0}$ is larger than $A^{CP}_{K^-\pi^+}$. In
SUSY models, the gluino contribution leads to a large value of
$\theta_P$ and depending on the sign of this angle the parameter
$r_T$ could be enhanced or reduced, see Eq.(\ref{risusy}). As will
be seen below, in this case we can explain the CP asymmetry
results with moderate values of the electroweak penguin parameter
$r_{EW}^C$. Note that in other models studied in the literatures,
the value of this parameter is required to be larger than one in
order to account for the CP asymmetry results.

Now let us discuss the SUSY contribution to the CP asymmetries
$A^{CP}_{K\pi}$. As can be seen from Table 1 that the experimental
measurements of $A^{CP}_{K^0\pi^0}$ suffer from a large uncertainty.
It turns out that it is very easy to have the SUSY results for this
asymmetry within the range of $2 \sigma$ measurements. Thus, this
decay mode is not useful in constraining the SUSY parameter space
and can be ignored in our discussion for the correlation among the
CP asymmetries of $B\to K \pi$ in generic SUSY models.

We will consider, as in the previous section, three scenarios with
a single mass insertion, two mass insertions, and three mass
insertions. In the first case, if we consider the contribution due
to the mass insertion $(\delta^u_{LR})_{32}$ the maximum values of
$\{k,l,m\}$ are given by $\{0.061,1.11,1.62\}$. While from
$(\delta^d_{LR})_{23}$ and $(\delta^u_{LL})_{32}$ one finds that
the maximum values of $\{k,l,m\}$ are $\{0.18,0.014,0.13\}$ and
$\{0.0019,0.053,0.13\}$ respectively. Note that $k$ is almost
negligible in the case of dominant chargino contribution which
depends on $(\delta^u_{LL})_{32}$ and $(\delta^u_{LR})_{32}$ and
can be significantly enhanced by the gluino contribution that
depends on $(\delta^d_{LR})_{23}$ as emphasized in
Ref.\cite{Gabrielli:2004yi}. Also from Eqs.(\ref{rEW},\ref{rEWC}),
and (\ref{risusy}), one finds \bea r_{EW} &=&
r_{EW}^{SM} (1+ l^2 + 2 l \cos \theta'_{EW})^{1/2},\\
r^c_{EW} &=& (r^C_{EW})^{SM} (1+ m^2 + 2 m \cos \theta^{C'}_{EW})^{1/2},\\
r_T &^=& \frac{r_T}{\vert 1+ k e^{i \theta'_P}\vert}. \eea Since
$(r^C_{EW})^{SM} \simeq 0.01$, the enhancement of $r^C_{EW}$ remains
quite limited in SUSY models and it is impossible to enhance it to
be of order one. Hence, the contribution of $r^C_{EW}$ to
$A^{CP}_{K^-\pi^+}$ is negligible respect to the contribution of
$r_{EW}$ to $A^{CP}_{K^-\pi^0}$. To overcome this problem and get
the desired relation between $A^{CP}_{K^-\pi^+}$ and
$A^{CP}_{K^-\pi^0}$ a kind of cancelation between $r_T$ and $r_{EW}$
contributions to $A^{CP}_{K^-\pi^0}$ is required. Such cancelation
can be obtained naturally without fine tuning the parameters if $r_T
\sim r_{EW}$, i.e. the total value of $r_T < r_T^{SM}$. This could
happen if $k$ is not very small. Therefore, one would expect that
the scenarios with dominant chargino contribution, where $k=0.061$
or $k=0.0019$ will not be able to saturate the experimental results
of $A^{CP}_{K^-\pi^+}$ and $A^{CP}_{K^-\pi^0}$ simultaneously. This
observation is confirmed in Fig. 3(top-left), where the results of
$A^{CP}_{K^-\pi^+}$ is potted versus the results of
$A^{CP}_{K^-\pi^0}$ for $\{k,l,m\}= \{0.061,1.11,2.62\}$ and the
other parameters vary as follows: $\delta_i \equiv -\pi,\pi/2$, and
$\pi$. The angles $\theta_{EW}$ and $\theta_{EW}^C \in [-\pi,\pi]$.
Also $\theta_P$ is assumed to be in the region $[\pi/4, \pi/2]$.
Note that in this plot we have taken the $A^{CP}_{K^0\pi^-}$ as
constraint. Thus all the points in the plot correspond to consistent
values of $A^{CP}_{K^0\pi^-}$ with the experimental results.


\begin{figure}[tpb]
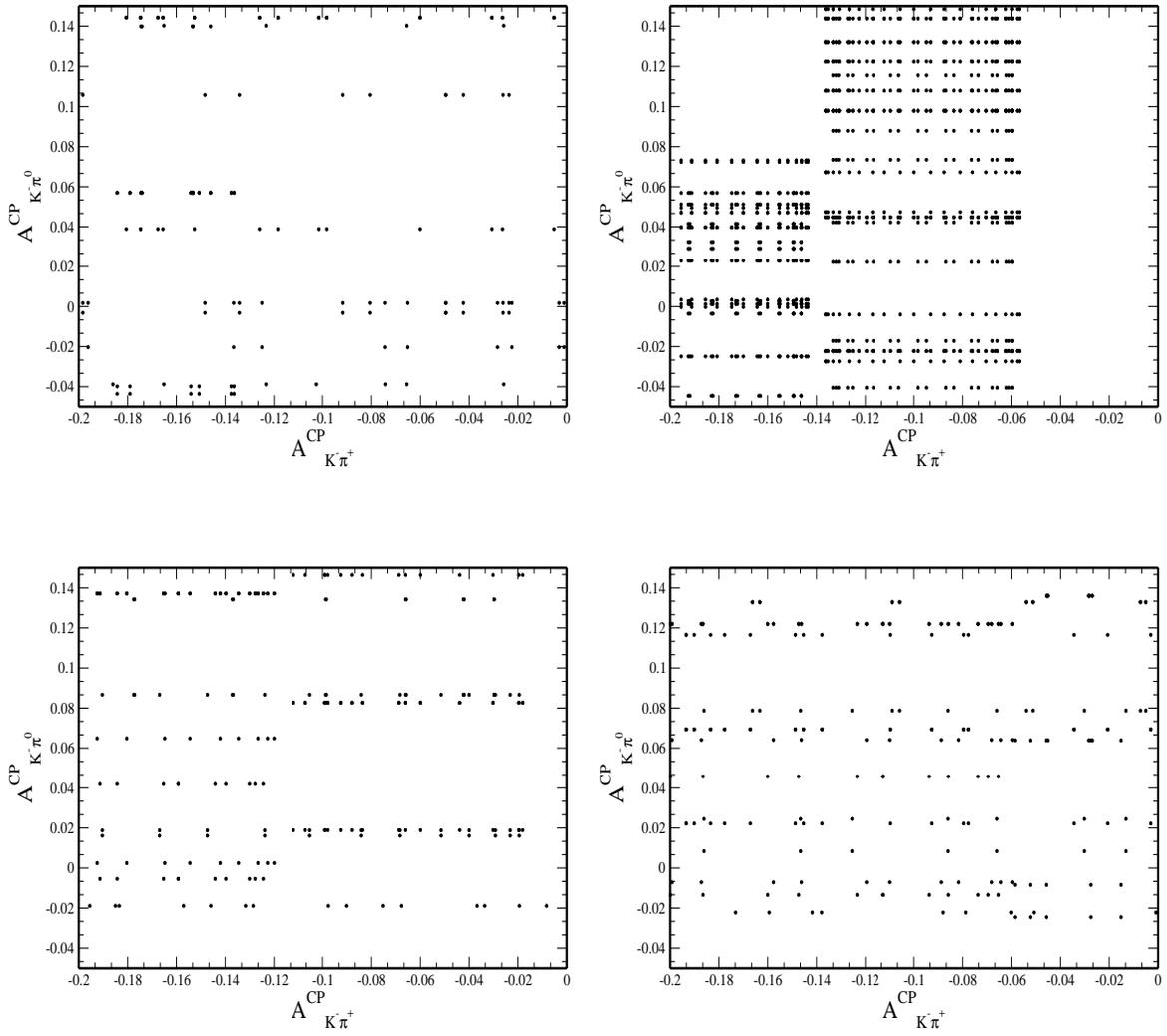

\begin{center}
\dofourfigs{3.1in}{asym1.eps}{asym2.eps}{asym3.eps}{asym4.eps}
\end{center}
\caption{\small CP asymmetry of $B\to K^- \pi^+$ versus CP
asymmetry of $B\to K^- \pi^0$ for $\{k,l,m\}= \{0.061,1.11,2.62\},
\{0.18,0.014,0.13\},\{0.24,1.12,1.48\},\{0.32,0.95,2.26\}$
respectively from left to the right and top to bottom. Strong
phases $\delta_i \equiv -\pi,\pi/2, \pi$ and CP violating phases
$\theta_{EW}$ and $\theta_{EW}^C$ reside between $-\pi$ and $\pi$.
Finally $\theta_P$ is assumed to be in the region $[\pi/4,
\pi/2]$.} \label{Fig3}
\end{figure}

Now we consider the second scenario with dominant gluino
contribution, {\it i.e.}, $(\delta^d_{LR})_{23}\simeq 0.005 e^{i
\pi/3}, (\delta^u_{LL})_{32}=(\delta^u_{RL})_{32}=0$. In this case,
one finds that the maximum values of $\{k,l,m\}$ are give by
$\{k,l,m\}=\{0.18, 0.014,0.13\}$, hence $r_T$ is reduced from
$r_T^{SM} \simeq 0.2$ to $r_T\simeq 0.12$, while $r_{EW}$ and
$r_{EW}^C$ approximately remain the same as in the SM. In Fig.
3(top-right) we plot the CP asymmetries $A^{CP}_{K^- \pi^+}$ and
$A^{CP}_{K^- \pi^0}$ in this scenario, with varying the relevant
parameter as before. It is remarkable that large number of points of
the parameter space can simultaneously accommodate the experimental
results of these CP asymmetries. It is slightly surprising to get
the values of the CP asymmetries $A^{CP}_{K\pi}$ within the
experimental range, {\it i.e.}, $A^{CP}_{K^-\pi^+} \in [-0.075,
-0.151]$ and $A^{CP}_{K^-\pi^0} \in [-0.04, 0.12]$, by just one mass
insertion in dominant gluino models. This is contrary to the
$R_c-R_n$ results which need gluino and chargino combination in
order to be within the experimental range. This result can be
explained by the cancelation that occurs in $A^{CP}_{K^-\pi^0}$
between the $r_T$ and $r_{EW}$ contributions and the negligible
effect of $r_{EW}^C$ to $A^{CP}_{K^-\pi^+}$.

To be more quantitative, let us consider the following example
where $(\delta^{d}_{LR})_{23} \simeq 0.005~ e^{i\pi/3}$ and
$(\delta^{u}_{LL})_{32}=(\delta^u_{RL})_{32}=0$. In this case one
get $r_T=0.12$, $r_{EW}=0.13$ and $r_{EW}^C = 0.01$. Therefore,
the main contribution to $A^{CP}_{K^-\pi^+}$ is due to the linear
term in $r_T$ which is $r_T \sin \delta_T \sin(\theta_P+\gamma)$.
With $\theta_P \sim \pi/3$ and $\delta_T \sim - \pi/4$, this
contribution leads to $A^{CP}_{K^-\pi^+} \simeq -0.113$. Since
$r_T$ gives the same contributions to $A^{CP}_{K^-\pi^0}$ a
significant positive contribution from $r_{EW}$ is required to
change the $A^{CP}_{K^-\pi^0}$ and make it positive. With $r_{EW}
= 0.13$, the $A^{CP}_{K^-\pi^0}$ is approximately given by
$A^{CP}_{K^-\pi^0} \simeq -0.113 + 0.26 \sin \delta_{EW}
\sin(\theta_P - \theta_{EW})$. It is worth mentioning that
although $\theta_P'$ and $\theta'_{EW}$ are equal in case of
single mass insertion, the value of $\theta_P$ and $\theta_{EW}$
are different due to the different value of $k$ and $l$. In this
example, it turns out that $\theta_P -\theta_{EW} \sim \pi/9$.
Hence, one gets $A^{CP}_{K^-\pi^0} \simeq -0.113 + 0.22 \sin
\delta_{EW}$. So that for $\delta_{EW} \sim \pi/4$, one finds
$A^{CP}_{K^-\pi^0} \simeq 0.04$ which is the central value of the
experimental measurements reported in Table 1.

We turn to the contributions from two mass insertions:
$(\delta^d_{LR})_{23}$ and $(\delta^u_{RL})_{32}$, which reflect
simultaneous contributions from the penguin diagrams with chargino
and gluino in the loop. Applying the $b \to s \gamma$ constraints
on these mass insertions, the maximum values of $\{k,l,m\}$ is
found to be $\{0.24, 1.12,1.148\}$. In this case we obtain
$r_T=0.11$, $r_{EW}=0.54$ and $r^C_{EW}=0.06$. Therefore, the CP
asymmetry $A^{CP}_{K^-\pi^0}$ is dominated by $r_{EW}$
contribution and in order to get $A^{CP}_{K^-\pi^0}$ of order
$\mathcal{O}(0.04)$, a small value of the strong phase
$\delta_{EW}$ should be used. This makes the possibility of
saturating the results of $A^{CP}_{K^-\pi^+}$ and
$A^{CP}_{K^-\pi^0}$ is less possible than the previous case. In
Fig. 3(bottom-left), we present the results of this scenario for
the same set of input parameters used before. This figure confirms
our expectation and it can be easily seen that it has less points
of the parameter space that account for the experimental results
of the CP asymmetries than Fig. 3(top-right). Note also that with
two mass insertions, the phases $\theta'_P$ and $\theta'_{EW}$ can
be considered independent, hence the angles $\theta_P$ and
$\theta_{EW}$ are also independent.

Finally we consider the case of three non-vanishing mass
insertions: $(\delta^d_{LR})_{23}$, $(\delta^u_{RL})_{32}$ and
$(\delta^u_{LL})_{32}$. Including the $b\to s \gamma$ constraints
we find that maximum value of $\{k,l,m\}$ is found to be $\{0.32,
0.95,2.26\}$. The corresponding values of $r_i$ are $r_T=0.10$,
$r_{EW}=0.48$ and $r_{EW}^C = 0.09$. It is clear that $r_T$ and
$r_{EW}$ are slightly changed than the previous scenario, while
$r_{EW}^C$ is enhanced a bit. In this case, it will be easier to
accommodate for $A^{CP}_{K^-\pi^0}$. The numerical results for
this scenario are given in Fig. 3(bottom-right) for the same set
of parameter space used in previous cases. As can be seen from
this figure, the probability of accommodating the experimental
results of different CP asymmetries in this class of models is
higher than it in models with two mass insertions. However, it
remains that the model with dominated gluino contributions
provides the largest possibility of saturating the experimental
results of CP asymmetries of $B\to K \pi$.
\subsection{SUSY contributions to the mixing CP asymmetry in $B \to
K^0 \pi^0$} We turn our attention, now, to the mixing CP asymmetry
of $B \to K^0 \pi^0$. As mentioned before, this decay is dominated
by $b\to s$ penguin. Thus, within the SM, the CP asymmetry $S_{K^0
\pi^0}$ should be very close to the value of $\sin 2\beta \simeq
0.73$. However, the current experimental measurements summarized
in Table 1 show that $S_{K^0 \pi^0}$ is lower than the expected
value of $\sin 2 \beta$, namely \be S_{K^0\pi^0} \simeq 0.34 \pm
0.28. \ee In this section we aim to interpret this discrepancy in
terms of supersymmetry contributions. It is useful to parameterize
the SUSY effects by introducing the ratio of SM and SUSY
amplitudes as follows \bea
\left(\frac{A^{\susy}}{A^{\sm}}\right)_{K \pi}\equiv R_{\pi}~
e^{i\theta_{\pi}}~e^{i\delta_{\pi}}, \label{ratioPI} \eea where
$R_{\pi}$ stands for the absolute value of $\left
\vert\frac{A^{\susy}(B\to K^0\pi^0)}{A^{\sm}(B\to
K^0\pi^0)}\right\vert$ and the angle $\theta_{\pi}$ is the SUSY CP
violating phase. The strong (CP conserving) phase $\delta_{\pi}$
is defined by $\delta_{\pi}=\delta^{SM}_{\pi}-
\delta^{SUSY}_{\pi}$. This parametrization is analogously for
those of $S_{K \phi}$ and $S_{K \eta^{\prime}}$
\cite{Khalil:2003bi,Gabrielli:2004yi}.  Using this
parametrization, one finds that the mixing CP asymmetry $S_{K^0
\pi^0}$ in Eq.(\ref{asym_pi}) takes the following form \bea
S_{K^0\pi^0}&=&\Frac{\sin 2 \beta +2 R_{\pi} \cos \delta_{\pi}
\sin(\theta_{\pi}+2 \beta)+ R_{\pi}^2 \sin (2 \theta_{\pi}+2
\beta)}{1+ 2 R_{\pi} \cos \delta_{\pi} \cos\theta_{\pi}
+R_{\pi}^2}. \label{cpmixing_pi} \eea

Assuming that the SUSY contribution to the amplitude is smaller
than the SM one {\it i.e.} $R_{\pi} \ll 1$, one can simplify the
above expressions as: \bea S_{K^0\pi^0}= \sin 2\beta +2\cos
2\beta\sin\theta_{\pi} \cos\delta_{\pi} R_{\pi} +{\cal
O}(R_{\pi}^2)\, . \eea In order to reduce $S_{K^0\pi^0}$ smaller
than $\sin 2 \beta$, the relative sign of $\sin\theta_{\pi}$ and
$\cos\delta_{\pi}$ has to be negative. If one assumes that
$\sin\theta_{\pi}\cos\delta_{\pi} \simeq -1$, then $R_{\pi} \gsim
0.2$ is required in order to get $S_{K^0 \pi^0}$ within $1\sigma$
of the experimental range.

In the QCDF approach, the decay amplitude of $B \to K^0 \pi^0$ is
given by Eq.(\ref{ampl4}). As in the case of $B\to \phi(\eta') K$
\cite{Gabrielli:2004yi}, we will provide the numerical
parametrization of this amplitude in terms of the Wilson
coefficients $\bf{C_i}$ and $\bf{\tilde{C}_i}$ defined according to
the parametrization of the effective Hamiltonian in Eq.(\ref{Heff})
\begin{equation}
H^{\Delta B=1}_{\rm eff}=\frac{G_F}{\sqrt{2}}\sum_i \left\{ {\bf
C}_{ i} Q_i \, +\, {\bf \tilde{C}}_{i} \tilde{Q}_i\right\} ,
\label{Heff_NP}
\end{equation}
where the operators basis $Q_i$ and $\tilde{Q}_i$ are the same
ones of Eq.(\ref{Heff}). By fixing the hadronic parameters with
their center values as in Table 1 of Ref.\cite{BN}, we obtain
\begin{equation}
A(B\to K^0 \pi^0) = - i \frac{G_F}{\sqrt{2}} m_B^2 F_+^{B\to K}
f_{\pi} \sum_{i=1..10,7\gamma,8g} H_i(\pi) ({\bf C}_i - {\bf
\tilde{C}}_i), \label{Api}
\end{equation}
where
\begin{eqnarray}
H_1(\pi)&\simeq & - 0.7 + 0.0003 i, \nonumber\\
H_2(\pi) &\simeq& - 0.21 + 0.037 i - 0.006 X_H,\nonumber\\
H_3(\pi) &\simeq& 0.22 - 0.076 i + 0.0045 X_A + 0.0003 X_A^2 +
0.0065 X_H ,\nonumber\\
H_4(\pi) &\simeq& 0.68 - 0.078 i ,\nonumber\\
H_5(\pi) &\simeq& 0.2 - 0.001 X_A + 0.004 X_A^2\nonumber\\
H_6(\pi) &\simeq& 0.68 - 0.078 i - 0.007 X_A + 0.014 X_A^2 ,\nonumber\\
H_7(\pi) &\simeq& 0.95 + 0.0004 X_A - 0.0014 X_A^2,\nonumber\\
H_8(\pi) &\simeq& -0.068 + 0.08 i + 0.002 X_A - 0.0047 X_A^2 -
0.009 X_H ,\nonumber\\
H_9(\pi) &\simeq&  -1.16 + 0.026 i - 0.0015 X_A - 0.0001 X_A^2 -
0.003 X_H ,\nonumber\\
H_{10}(\pi) &\simeq&  - 0.67 + 0.08 i - 0.0096 X_H ,\nonumber\\
H_{7\gamma}(\pi) &\simeq&  0.0004,\nonumber\\
H_{8g}(\pi) &\simeq& - 0.045. \label{H_eta}
\end{eqnarray}
The different sign between ${\bf C}_i$ and ${\bf \tilde{C}}_i$
appearing in Eq.(\ref{Api}) is due to the fact that $\langle K^0
\pi^0 \vert Q_i \vert B \rangle = - \langle K^0\pi^0 \vert
\tilde{Q}_i \vert B \rangle,$ since the initial and the final
states have different parity. Comparing the coefficients
$H_i(\pi)$ with $H_i(\phi)$ and $H_i(\eta^{\prime})$ in
\cite{Gabrielli:2004yi}, one finds that the Wilson coefficients in
these decay amplitudes are different. Thus it is naturally to have
different CP asymmetries $S_{K^0 \pi^0}$, $S_{K \phi}$ and $S_{K
\eta'}$, unlike the SM prediction.

In order to understand the dominant SUSY contribution to the CP
asymmetry $S_{K^0\pi^0}$, it is useful to present a numerical
parametrization of the ratio of the amplitude $R_{\pi}$ in terms
of the relevant mass insertions. For the usual SUSY configurations
that we have used in the previous sections, we obtain \bea
\hspace{-0.5cm}R_{\pi}&\simeq& \left\{0.02 \times e^{-i 0.4}
 \dd{LL}{23} - 40.4 \times e^{-i 0.01} \dd{LR}{23}\right\} -
 \left\{ L \leftrightarrow R \right\} \no
 &+& 0.15\times e^{-i 0.002} \du{LL}{32} -  0.08 \times e^{-i 0.013}
 \du{RL}{32}.
\eea
From this result, it is clear that the largest SUSY effect is
provided by the gluino contribution to the chromo-magnetic
operator which is proportional to $\dd{LR}{23}$ and $\dd{RL}{23}$.
For $\dd{LR}{23}\simeq 0.006 \times e^{i \pi/3}$ and all the other
mass insertions set to zero, one finds $S_{K^0\pi^0} \simeq 0.34$
which coincides with the central value of the experimental results
reported in Table 1. It is important to note that with such value
of $\dd{LR}{23}$ the gluino contribution can account for the CP
asymmetries $S_{K\phi}$ and $S_{K\eta'}$ as well
\cite{Gabrielli:2004yi}. Furthermore, if we consider the scenario
where both chargino and gluino exchanges are contributed
simultaneously, the result of $R_{\pi}$ is enhanced and we can get
smaller values of $S_{K^0\pi^0}$.
\section{Conclusions}
In this paper we have analyzed the supersymmetric contributions to
the direct and mixing CP asymmetries and also to the branching
ratios of the $B\to K \pi$ decays in a model independent way.

We have shown that, in the SM, the $R_c-R_n$ puzzle which reflects
the discrepancy between the experimental measurements of the
branching ratios and their expected results can not be resolved.
Also the direct CP asymmetries $A^{CP}_{K^0\pi^-}$ and
$A^{CP}_{K^0\pi^0}$ are very small while $A^{CP}_{K^0\pi^0}$ and
$A^{CP}_{K^-\pi^+}$ are of the same order and can be larger. These
correlations among the CP asymmetries are inconsistent with the
recent measurements. Moreover the mixing CP asymmetry
$S_{K^0\pi^0}$, which is expected to be $\sin 2\beta$, differs from
the corresponding experimental data. The confirmation of these
discrepancies will be a clear signal for new physics beyond the SM.

We have emphasized that the $Z$-penguin diagram with chargino in
the loop and the chargino electromagnetic penguin can enhance the
contribution of the electroweak penguin to $B\to K \pi$ which is
supposed to play a crucial role in explaining the above mentioned
discrepancies. We, however, found that these contributions alone
are not enough to solve the $R_c-R_n$ puzzle. It turns out that a
combination of gluino and chargino contributions is necessary to
account for the results of $R_c$ and $R_n$ within the $b\to
s\gamma$ constraints. Nevertheless, our numerical results
confirmed that the general trend of SUSY models favors that the
experimental result of $R_c$ goes down.

We have also provided a systematic study of the SUSY contributions
to the direct CP asymmetries for $B\to K \pi$ decays. We found
that a large gluino contribution is essential to explain the
recent experimental data. It is worth mentioning that a large
gluino contribution is also important to accommodate another
controversial results measured in the B factories, namely the
mixing CP asymmetries $S_{\phi K}$ and $S_{\eta' K}$. Unlike the
$R_c-R_n$ puzzle, we found that the CP asymmetries $A^{CP}_{K\pi}$
can be saturated by a single mass insertion $(\delta^d_{LR})_{23}$
contribution. It has been noticed that a large electroweak penguin
is less favored by the CP asymmetries $A^{CP}_{K\pi}$. Therefore,
one needs to optimize the gluino and chargino contributions in
order to satisfy simultaneously the branching ratios and the CP
asymmetries of $B\to K\pi$.

Finally we have considered the mixing CP asymmetry $S_{K^0\pi^0}$.
We found, as in $S_{\phi K}$ and $S_{\eta' K}$, that the gluino
contribution through the LR or RL mass insertion gives the largest
contribution to $S_{K^0\pi^0}$. On the other hand, it is quite
possible for the gluino exchanges to account for $S_{K^0\pi^0}$,
$S_{\phi K}$ and $S_{\eta' K}$ at the same time.

\vspace{0.7in}

\noindent {\bf Acknowledgments}

\vspace{0.15in} \noindent I would like to thank E. Kou for very
useful discussions and fruitful collaboration in analyzing the
effect of SUSY contributions to $R_c$ and $R_n$. Also I would like
to thank A. Arhrib for enlightening conservations and interesting
comments.


\end{document}